\documentclass[prc,preprint,showpacs,showkeys,footinbib]{revtex4}
\usepackage{graphicx} 
\usepackage{amssymb} 
\usepackage{amsmath}
\def\bea{\begin{eqnarray}} 
\def\eea{\end{eqnarray}}

\begin{document}
\title{\hskip4in   \\ Coherent Vector  Meson Photo-Production from
Deuterium at Intermediate Energies}
\author{T.C. Rogers$^a$\email[E-mail:]{rogers@phys.psu.edu},
M.M. Sargsian$^b$, M.I. Strikman$^a$}
\email[E-mail:]{rogers@phys.psu.edu} \affiliation{$^a$Department of
Physics, Pennsylvania State University,\\ University Park, PA  16802,
USA\\ $^b$Department of Physics, Florida International University,
Miami, FL 33199 } \date{\today{}}

\begin{abstract}
We analyze the cross section for vector meson  photo-production off a
deuteron for the intermediate range of photon energies starting at 
a few GeVs above the threshold and higher.   We reproduce the steps in the derivation of the
conventional non-relativistic  Glauber expression based  on an
effective diagrammatic method while making corrections for
Fermi motion and intermediate energy kinematic effects.  We show that,
for intermediate energy vector  meson production, the usual
Glauber factorization breaks down and we  derive corrections to the
usual Glauber method to linear order in  longitudinal nucleon
momentum.  The purpose of our analysis is to establish  methods for
probing interesting physics in the production mechanism  for
$\phi$-mesons and heavier vector mesons.  We demonstrate how neglecting the breakdown of Glauber
factorization can lead to errors in measurements of basic cross sections
extracted from nuclear data.
\end{abstract}

\maketitle 

\section{Introduction}
Coherent vector meson production from nuclei has proven to be a useful
tool for  studying the structure of vector mesons.   In the very high
energy, small angle scattering regime, well above the threshold for
vector meson production, the large volume of available experimental
data involving proton targets  consistently supports the validity of
the vector meson dominance~(VMD) model for small photon
virtualities~\cite{Bauer:1977iq,feynman}.  This, combined with the onset of
the eikonal regime in the diffractive region  has lead to the
development of a successful  theoretical framework for the description
of vector meson photo-production  off nuclei based on the combined  VMD
model and Glauber theory of hadron-nuclei
rescattering~\cite{Bauer:1977iq,Franco:1965wi}.  The simplicity of the VMD-Glauber
framework arises from   the fact that at high energies the basic
$\gamma N\rightarrow V N$ and $V N\rightarrow N N$ amplitudes  vary
slowly with the total energy of the $\gamma N$ system  relative to the
range of important energies in the deuteron wavefunction.   This
observation leads to the factorizability of  the basic $\gamma N
\rightarrow V N$ amplitude  from the momentum space integral, and
yields the conventional  Glauber multiple scattering series consisting
of non-relativistic form factors and elementary scattering amplitudes.

The VMD-Glauber theory has lead, in particular, to the demonstration
that the coherent photo-production of vector mesons off the deuteron at
large $-t$ is defined mainly by the rescattering  contribution
\cite{SLACexp}.  Since the $VN\rightarrow V N$ amplitude  appears in
the double scattering term, one may use  nuclear photo-production
reactions to study the properties of vector mesons~\cite{VM2}.  By
choosing different $t$, one can control the relative  distance at
which rescattering may occur, which allows one to     investigate the
space-time evolution of hadronic systems produced in electro(photo)-
production.

The above program can be extended to the study of coherent vector
electro-production at large $Q^2$. In this case,  coherent vector
meson production from the deuteron can be used to   study color
coherence/transparency phenomena in vector  meson electro-production at
high $Q^2$.  The onset of color transparency will reveal itself
through the substantial drop in the double scattering contribution
with an increase of $Q^2$ as opposed  to the nearly energy independent
behavior of the double scattering term for the  generalized VMD
prediction~\cite{VM2}.

In this paper we consider yet another venue of application for vector
meson  photo-production off nuclear targets by considering
photo-production in the intermediate range of energies starting a few GeVs above the
threshold.  These reactions have great potential for probing several
effects such as non-diffractive, OZI violating mechanisms for vector
meson production  mesons, in-medium modifications of vector mesons,
the importance of ``non-ideal'' $\omega-\phi$ mixing, and other new
mechanisms for vector meson production (see
Refs.~\cite{Titov:1999eu,Titov:2002zy,Oh:2003gm}).

Finally, it would be interesting to learn whether the $\phi$-meson is
produced with a small enough transverse size that quark degrees of
freedom may become relevant, as in the case of $J/\psi$-production.
Actually, in the  case of $J/\psi$, the cross section of the
$J/\psi-N$ interaction $\sigma_{J/\psi-N}\sim 3 \, mb$,
\cite{Frankfurt:1985cv} estimated based on the A-dependence of $J/\psi$
photo-production at energies $\sim 20 $~GeV, is much larger than the
estimate based on the VDM:  $\lesssim 1 \, mb$.  This is likely due to the
color transparency phenomenon~\cite{Frankfurt:1985cv}.  
A natural question is whether a trace of this effect remains in the case 
of $\phi$-production.  Jefferson Lab has 
produced data for $\phi$ production that is currently being analyzed.

The interest in intermediate energy reactions makes  it necessary to
re-evaluate  the assumptions of the traditional Glauber series method,
and to develop a new theoretical approach.  This paper addresses  the
issues one must face when considering photon energies large  enough 
that the eikonal approximation is an appropriate description of  hadronic re-interactions, but not large enough that it
is appropriate to neglect vector meson  masses in kinematical
calculations or any non-trivial  s-dependence of  the amplitude for
photo-production of vector mesons from the nucleon.  Furthermore, for
small photon energies ($\lesssim 3$~GeV) the VMD hypothesis becomes
suspect as a description of the $\gamma N \rightarrow V N$ amplitude.  
Therefore, we will not restrict ourselves to VMD model of 
$\gamma N \rightarrow V N$ amplitude, considering instead  
the adequately parameterized form of photon-nucleon amplitudes. 
We argue in this paper that there may be a range of photon energies for
which the eikonal approximation is valid, but where the usual Glauber theory
assumptions of factorization and ultra-relativistic kinematics break down.

Although we retain the eikonal approximation, our approach is
distinctly different from the usual Glauber-VMD approach.  In
particular, one of  the basic assumptions used in the Glauber approach
is that  the basic $\gamma N\rightarrow V N$ and $V N\rightarrow V N$
cross sections are slowly varying functions of  center of mass energy
and that the small Fermi momentum of the  nucleons can be neglected in
the evaluation of the total center of mass energy of the $\gamma N$ and $VN$ systems.
These assumptions  result in the usual factorizability already
discussed above.  At intermediate energies, however, the photon
energy is comparable to the vector meson mass, and the basic amplitude
may gain non-trivial energy dependence due to the fact that Regge theory 
may be inadequate at intermediate photon energies.  
The  usual smooth, slow rise in the total $\gamma N
\rightarrow V N$  cross section characteristic of high energy
diffractive scattering may be absent at intermediate energies.  Fermi motion
effects thereby destroy the factorizability  of nuclear scattering
into basic amplitudes and form factors.  
Also, the longitudinal
momentum transfered (proportional to $M_V^2/E_{\gamma}$) plays an
important role as compared  to reactions in the diffractive regime and
further calls into  question the factorization assumption.  Earlier work (e.g.~\cite{Kolbig:1968rm}) 
has considered the effect of longitudinal momentum transfers, but the breakdown of factorization has 
not been discussed.

To summarize, the particular reaction we are interested in in this paper
is the coherent photoproduction of vector mesons from the 
deuteron.  However, the energy dependence of the $\gamma N \rightarrow V N$
will require that we
account for Fermi motion effects which, in turn, will require that we
account for non-factorization effects.   In the derivation of the
total $\gamma D \rightarrow V D$ amplitude we  will use the Generalized
Eikonal Approximation (GEA) with effective Feynman diagram rules (see, e.g.~\cite{Gribov:1968jf,Bertocchi,edepn,GEA}). 
This is the approximation, valid at
appropriately high energies, that allows us to derive the
scattering amplitudes starting with corresponding effective Feynman
diagrams while neglecting multiple scattering from the same nucleon.  
This is very similar to the GEA approach that has been applied to the $A(e,e^{\prime} p)X$ reactions on the
nucleus~\cite{GEA}. 

By maintaining the result in terms of momentum space integrals, within the GEA,
transferred longitudinal momentum and Fermi motion effects may be explicitly taken into
account consistently.  In our derivations,  we  keep only   the
corrections to the basic amplitudes that are of linear order in
longitudinal exchanged momentum    or nucleon momentum    (neglecting
order ${\bf k_{N}}^2/m_{N}^2$ corrections, where ${\bf k_{N}}$  is the bound state
nucleon momentum). This  allows us to relate the $D\rightarrow NN$
transition vertex to the nonrelativistic wavefunction of the
deuteron.  Since dynamical, model-dependent corrections related to the N-N interaction
are expected to be of quadratic or higher order in nucleon momentum, then 
linear order corrections arising from intermediate energy kinematics should be
taken into account \emph{before} any specific theory of the basic bound state amplitude
that deviates from the nearly flat behavior of Regge theory is considered and used in the typical Glauber theory approach.

As it was explained above, we work in the kinematic regime in which 
diffractive behavior is not yet fully established
but the momenta of the produced vector mesons are high enough that the eikonal
approximation for the hadronic rescatterings is justified.
As a result, a formalism should be maintained that 
allows the $\gamma N \rightarrow V N$ and the $V N \rightarrow V N$ amplitudes
to be independently modeled.  Fitting data to
our modified form of the Glauber theory by using the $V N
\rightarrow V N$  amplitude as a parameter allows one to infer a
value for the $V N \rightarrow V N$ cross section.  We emphasize that the main steps of
this paper have been known for several decades; the Glauber theory in terms of 
effective Feynman diagrams was established in Ref.~\cite{Gribov:1968jf}.  The
effects of longitudinal momentum transfer in terms of 
phase shifts have also been studied~\cite{Kolbig:1968rm}.  However, 
as far as we are aware, there has never been direct numerical study 
of the effect of the breakdown of factorization in Glauber theory as it applies  
vector meson production.  (The effects of factorization break-down in proton knock-out have
been studied in Ref.~\cite{Jeschonnek:2000nh}.)  One result that we find is
that the breakdown of factorization persists even in the limit that off-shell effects in 
the bound state nucleon amplitudes are negligible.  

The paper is organized as follows: In Sec.~\ref{sec:calculation} we derive scattering
amplitudes  for the $\gamma D\rightarrow VD$ reaction based on the
generalized eikonal approximation.  In Sec.~\ref{sec:analysis} we discuss 
the steps needed to take into account linear order corrections in nucleon momentum.  
In Sec.~\ref{section4} we perform a 
sample calculation where we compare our results
at intermediate energy kinematics with the prediction of conventional
VMD-Glauber theory. We identify the effects which are 
responsible for the divergence of the our approach from the standard Glauber
theory.  We demonstrate that  effects calculated in this paper, if
unaccounted for, can yield a misinterpretation of  the $V N$
scattering cross section if it is extracted from the data using the usual
Glauber approximation. 
In a related issue, we will discuss the recent data on $\phi$ production at SPring-8/LEPS~\cite{Mibe:2005er} and
demonstrate the need to consider kinematic effects in the intermediate energy region in Sec.~\ref{sec:future}.  In particular,
this data suggests the importance of non-vacuum exchanges corresponding to $\eta$ and $\pi$ in the $\phi$-meson production 
mechanism at $E_{\gamma}$ of a few GeVs. 
In Sec.~\ref{sec:summary} we summarize our results.  In 
Appendix~\ref{sec:app4} we describe the model of the basic amplitude that we used for
our sample calculation, and in Appendix~\ref{sec:app3} we give an overview of the usual treatment of 
deuteron spin in Glauber theory.

\section{Formulae for the Amplitudes}
\label{sec:calculation}

\subsection{Reaction and Kinematics}
\label{sec:born}

We study the coherent photo-production of vector mesons off the
deuteron in the reaction:
\begin{equation}
\gamma + D\rightarrow V + D^\prime,
\label{reaction}
\end{equation}
where $P\equiv (M_D, 0)$ and $P^\prime\equiv (E_D, {\bf P^{\prime}})$
define the initial and final four momenta of the deuteron.  We use
natural units ($c = \hbar = 1$).  
$q \equiv
(E_\gamma,{\bf q})$ and $P_V\equiv (E_V,{\bf  P_V})$ define the
4-momenta of the initial photon and the final state meson
respectively.  The three-momentum transfered is defined as ${\bf
l}={\bf q} - {\bf P_V}$.  

\begin{figure}
\centering \includegraphics [scale=0.90]{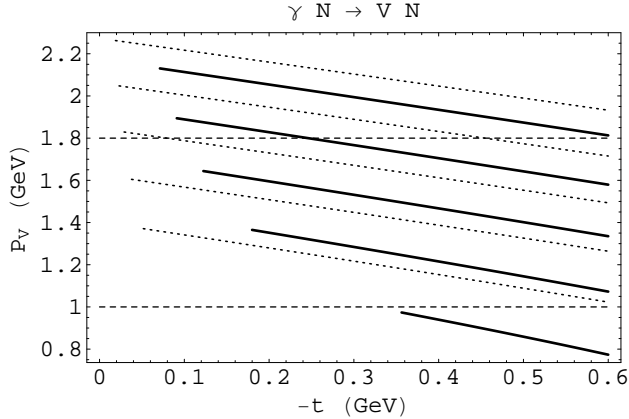}
\caption{3-momentum of a vector meson produced by a photon scattering off a nucleon target as a function of $-t$ for a given 
set of fixed $E_\gamma$.  The solid lines correspond to $\phi$-meson
production, whereas the dotted lines correspond to $\rho^{0}$-meson
production.  The incident photon energies in each case, going from the bottom
curve to the top curve are 1.6,1.8,2.0,2.2,2.4~GeV.  Details are discussed in the text.}
\label{fig:kinematics}
\end{figure}

In our calculations we concentrate on  \emph{intermediate} energy kinematics 
in which, although the photon energies are not high enough 
for the diffractive regime to be established for the photoproduction amplitude,
the produced vector meson is sufficiently energetic that the eikonal approximation 
can be applied to the calculation of final state hadronic rescatterings.  This require 
further elaboration:  
The GEA is the ``straight line'' approximation in that the
incident particle follows a nearly straight line path through the nucleus.  
Clearly this must occur at high enough energies that higher partial 
waves than just the $s$-wave contribute.
In order to establish the appropriate kinematical regime for our approach, we
have plotted in
Fig.~\ref{fig:kinematics} the lab frame 3-momentum of the
final state vector meson as a function of $-t$ for a set of  incident
photon energies for the case of
$\rho^{0}$ and $\phi$-meson production from a nucleon.   
Experience with the application of the Glauber model to the description of  
proton-nucleus scattering~\cite{Alkhazov:1978et} as well as $A(e,e'p)X$ 
reactions~\cite{Garrow} indicates that the
eikonal approximation works roughly for $p_{N}/m_{N} \gtrsim 1$ ($p_{N}$ is the proton 3-momentum) and it works  
extremely well for $E_{N} \gtrsim 2 m_{N}$.  Since $m_{\phi}\approx m_N$,  we expect to see the onset of the applicability of the GEA  
for a similar range of momenta for the case of $\phi$-meson production.
By analogy with the proton case, we continue to use the criterion that $E_{\gamma} \gtrsim 2 M_{V}$, and we find that the 
value of vector meson 3-momentum above which the GEA may certainly be applied is 
$P_{V} \gtrsim 1.8$~GeV for $\phi$-meson production.  The values 
of 3-momentum, $1$~GeV and $1.8$~GeV have been indicated by horizontal dashed lines in Fig.~\ref{fig:kinematics}.
These dashed lines in Fig.~\ref{fig:kinematics}
may be viewed as separating kinematic configurations where our approach may 
be applied to $\phi$-meson production from kinematic regions where both the approach of this paper and the standard Glauber approach should
be abandoned entirely with regards to $\phi$-meson production.  Below $P_{V} \approx 1$~GeV, both the approach of this paper and
the usual Glauber approach should be abandoned.  Between $1$~GeV and $1.8$~GeV, the GEA may become a rough approximation, but
above $1.8$~GeV, the approach that we take in this paper by using the GEA is a very good approximation.  For the $\rho^{0}$-meson, 
the eikonal regime begins at smaller values of momentum than for the $\phi$ meson due to its smaller mass, so to avoid confusion 
we do not include the corresponding range of applicability of the eikonal approach to $\rho^{0}$-meson production in Fig.~\ref{fig:kinematics}. 

\begin{figure}
\centering \includegraphics [scale=0.50]{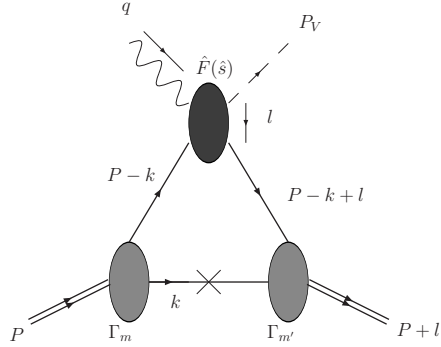}
\caption{The impulse diagram for photo-production.  The cross on the
spectator nucleon line indicates that the spectator nucleon will be
taken  on shell in the non-relativistic approximation.  (For all
Feynman graphs we use Jaxodraw~\cite{Binosi:2003yf}.)}
\label{fig:figure0}
\end{figure}

\begin{figure}
\centering \includegraphics [scale=0.50]{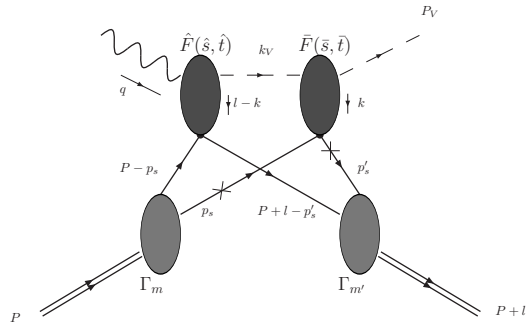}
\caption{The double scattering diagram for photo-production.  The crosses on the spectator
nucleon lines indicate that we will take poles corresponding to these nucleons going on-shell.  (See sec.~\ref{sec:double})}
\label{fig:figure2}
\end{figure}

Our main interest in this paper is the production of the $\phi$-meson
at around $3$~GeV, so the application of the GEA is quite safe.  We will find that another problem 
arises at $t \approx t_{min}$, and this will be discussed in Sec.~\ref{section4}, but the above 
argument remains applicable as long as $-t$ is more than a few tens of MeVs larger than $-t_{min}$.
We further assume that the non-relativistic  model of the $N-N$ interaction
can be represented by a $D \rightarrow N N$ vertex.  In the case of the
deuteron, there are only two relevant
diagrams: the single  scattering diagram (Born term) of
Fig.~\ref{fig:figure0} and the  double scattering diagram of
Fig.~\ref{fig:figure2}. 

\subsection{The Born Amplitude}
\label{sec:borncalc}
We start with the calculation of the amplitude corresponding to the
Born  term of Fig.~\ref{fig:figure0}.  $F^0_{m,m^\prime}(s,t)$ will
denote the $\gamma D \rightarrow V D$  scattering amplitude for the
Born term in which only one of the nucleons takes part in the
interaction,   whereas $\hat{F}(\hat{s},t)$ will denote the basic
$\gamma N\rightarrow VN$ scattering amplitude.  A hat on a  variable
indicates that it is associated with the $\gamma N \rightarrow V N$
subprocess rather than the process of Eq.~(\ref{reaction}).  The
superscript, $0$, is meant to distinguish the Born term from the double
scattering term.  The initial and final polarizations of the deuteron
are denoted by $m$  and $m^\prime$ respectively.  Because we consider 
only intermediate energies, the $\hat{F}(\hat{s},t)$ amplitude  is not
necessarily diffractive and we do not assume the validity of  the VMD
hypothesis.  We  neglect the spin-flip component of the basic
amplitude (i.e.  $\hat{F}$ and $\bar{F}$ are approximately diagonal is
nucleon spin.)  The $D \rightarrow NN$ vertex is denoted by
$\Gamma_{m}$.  All variables correspond to the labels in
the Feynman diagram of Fig.~\ref{fig:figure0} for the single
scattering (Born) term.  The free nucleon mass is denoted by $m_{N}$.
 
By applying effective Feynman rules to the
graph in Fig.~\ref{fig:figure0} we obtain the covariant
scattering amplitude,
\begin{equation}
\begin{split}
  & F^{0}_{m,m^{\prime}}(s,t)  = \\ & - \int \frac{d^{4} {\bf k}}{i(2
  \pi)^{4}} \frac{ \Gamma^{\dagger}_{m^{\prime}}(P - k +l)\hat{F}
  (\hat{s},t) \Gamma_{m}(P -k) }{\left[ (P-k+l)^{2} - m_{N}^{2} + i
  \epsilon \right] \left[ (P-k)^{2} - m_{N}^{2} + i \epsilon \right]
  \left[ k^{2} - m_{N}^{2} + i \epsilon \right] } \\  + & (p
  \leftrightarrow n). \label{eq:born1}
\end{split}
\end{equation}
$(p \leftrightarrow n)$ refers to the term in which the neutron and
proton are inverted.  In the remainder of this text, Mandelstam
variables that appear within an integral are understood to be
functions  of  internal nucleon 4-momentum and the incident photon
4-momentum.

We proceed with the derivation by estimating the loop integral in
Eq.(\ref{eq:born1}) up to terms of order $\frac{{\bf
k}^2}{m_N^2}$. This approximation allows us to evaluate the integral
in Eq.(\ref{eq:born1}) by keeping only the pole  contribution
which yields a positive energy for the spectator nucleon.  We find,
\begin{equation}
\begin{split}
  F^{0}_{m,m^{\prime}}(s,t) = \int \frac{d^{3} {\bf k}}{(2 \pi)^{3}}
  \frac{\Gamma_{m^{\prime}}^{\dagger}(P - k + l) \hat{F}(\hat{s},t)
  \Gamma_{m}  (P - k) }{ 2 k_{0}  \left[ (P-k+l)^{2} - m_{N}^{2} + i
  \epsilon \right]  \left[ (P-k)^{2} - m_{N}^{2} + i \epsilon \right]
  } \\ + (n \leftrightarrow p).  \label{eq:born4}
\end{split}
\end{equation}
(Note that $k_{0} = m_{N}$ up to correction terms of order ${\bf k}^{2}/m_{N}^{2}$.)
We now make use of the correspondence between the non-relativistic
wave function and the vertex function,
\begin{equation}
\tilde{\Psi}_{m} ({\bf k}_{rel}) \equiv \frac{ -\Gamma_{m}(P -
k)}{2\sqrt{k_{0} (2 \pi)^{3}}D(P - k)}, \label{eq:born5}
\end{equation}
the form of which is established by the Lippman-Schwinger Equation~\cite{Brown:1973xt}
and  by demanding that the non-relativistic wavefunction be normalized
to unity.   Here, $-D({P-k})$ is the propagator denominator of the
struck nucleon.   We write ${\bf k}_{rel}$ to indicate that the
argument of the wave function is the {\bf relative} 3-momentum of the
two nucleons.  Using Eq.~(\ref{eq:born4}) with Eq.~(\ref{eq:born5})
and using lab frame kinematics yields,
\begin{eqnarray}
F^{0}_{m,m^{\prime}}(E_{\gamma},l) &=&  2 \int d^{3} {\bf k}
\tilde{\Psi}_{m^{\prime}}^{\dagger} ( {\bf k} - {\bf l}/2 )
\hat{F}(E_{\gamma},k,l) \tilde{\Psi}_{m}({\bf k}) + (n \leftrightarrow
p). \label{eq:born6a}
\end{eqnarray}
We stress, at this point, that Eq.~(\ref{eq:born6a}) does {\bf not} yet
coincide with the conventional VMD-Glauber theory because we have
abandoned the usual assumptions that allow us to ignore the ${\bf
k}$-dependence in the basic amplitude, which would normally allow us
to factor the basic amplitude out of the integral and leave us with
the product of the basic amplitude with the non-relativistic form
factor.  For heavier vector mesons (like the $\phi$-meson), the
vector meson mass may not be negligible, and the $\hat{s}$-dependence
of the  basic amplitude becomes non-trivial at intermediate photon energies.

\subsection{The Double Scattering Amplitude}
\label{sec:double}

Having obtained the Born term in Eq.~(\ref{eq:born4}), we move on to
calculate the double scattering term of Fig.~\ref{fig:figure2}.
Applying the effective Feynman diagrammatic rules, we obtain
\begin{equation}
\begin{split}
  & F^{1}_{m,m^{\prime}}(s,t)  = \\ & - \int \frac{d^{4}p_{s}}{i(2 \pi)^{4}}
  \frac{d^{4}p^{\prime}_{s}}{i(2 \pi)^{4}}
  \frac{\Gamma_{m^{\prime}}^{\dagger}(P + l - p_{s}^{\prime} )
  \bar{F}(\bar{s},\bar{t})  \hat{F}( \hat{s},\hat{t} ) \Gamma_{m}(P -
  p_{s} )}{\left[ p_{s}^{2} - m_{N}^{2} + i \epsilon \right] \left[
  p^{\prime 2}_{s} - m_{N}^{2} + i \epsilon \right]  \left[ (P -
  p_{s})^{2} - m_{N}^{2} + i \epsilon \right]} \\  & \times 
  \frac{1}{\left[ (P + l - p^{\prime}_{s})^{2} - m_{N}^{2} + i
  \epsilon \right] \left[ (q - l + p^{\prime}_{s} - p_{s})^{2} -
  M_{V}^{2} + i \epsilon \right]}   + ( p \leftrightarrow n
  ). \label{eq:double0}
\end{split} 
\end{equation}
Figure~\ref{fig:figure2} and Eq.~(\ref{eq:double0}) express the
following sequence of events: The incident photon scatters from a
nucleon with center of  mass energy, $\sqrt{\hat{s}}$, producing an
intermediate state with invariant mass, $M_{V}$.  The intermediate
state propagates through the deuteron before scattering from the other
nucleon with center of mass energy $\bar{s}$.  (Bars over variables
will  indicate that they correspond to the secondary scattering.)
We neglect fluctuations of the intermediate state for the present 
purposes.  
Now let us integrate over, $p_{s,0}$  and $p^{\prime}_{s,0}$.
The integration over $p_{s,0}$ is similar to the integration over $k_0$
for the Born term of Eq.(\ref{eq:born1}).  For the
$p^{\prime}_{s,0}$ integration one can choose one of the positive
energy poles at
$M_D+l_0-\sqrt{m_N^2+({\bf l}-{\bf p^\prime_s})^2}+i\epsilon$ and
$\sqrt{m_N^2+{\bf p^\prime_{s}}^2}-i\epsilon$ at the upper and lower
complex semiplane of $p^\prime_{0}$. Note that within
the approximation in which $p^{\prime,2}_s/m_N^2$, $l^2/m_N^2$
terms are consistantly
neglected, the integration over either pole will yield the same 
result.
We choose the  $\sqrt{m_N^2+{\bf p^\prime_{s}}^2}-i\epsilon$ pole
(the poles chosen for integration are identified by the crosses shown in
Fig.~\ref{fig:figure2}) because this choice reproduces the usual Glauber
formula in the most direct way.
Applying the definition in Eq.~(\ref{eq:born5}),  we recover the
formula quoted in~\cite{VM2},
\begin{equation}
\begin{split}
  F^{1}_{m,m^{\prime}}(E_{\gamma},l) \\ & =  - \int \frac{d^{3}{\bf
  p_{s}}^{\prime} d^{3}  {\bf p}_{s} }{(2 \pi)^{3} }
  \frac{\tilde{\Psi}_{m^{\prime}}^{\dagger}(\frac{{\bf l}}{2} - {\bf
  p_{s}}^{\prime} )    \bar{F}(\bar{s},\bar{t})
  \hat{F}(\hat{s},\hat{t}) \tilde{\Psi}_{m}(-{\bf p}_{s}
  )}{\sqrt{p_{s,0}p^{\prime}_{s,0}}\left[ (q - l  + p^{\prime}_{s} -
  p_{s})^{2} - M_{V}^{2} + i \epsilon \right]}.
\label{eq:double6}
\end{split}
\end{equation}
The $(p \rightarrow n)$ term is implicit in these equations.  Finally,
we put this equation into a form that makes the next section slightly
more manageable by transforming the variables of integration from
$p^{\prime}_{s}$ and $p_{s}$ to $p \equiv (p_{s} + p_{s}^{\prime})/2$
and $k  \equiv p^{\prime}_{s} - p_{s}$, and we make the redefinitions,
$k \rightarrow k + l/2$, and $p \rightarrow p + \frac{l}{4}$.  The
result of these changes is:
\begin{equation}
\begin{split}
  F_{m,m^{\prime}}^{1}(E_{\gamma},l) \\  =& - \int \frac{d^{3}{\bf p}
  d^{3} {\bf k} }{(2 \pi)^{3} }
  \frac{\tilde{\Psi}_{m^{\prime}}^{\dagger}( {\bf p} + \frac{{\bf
  k}}{2})  \bar{F}(\bar{s},\bar{t}) \hat{F}(\hat{s},\hat{t})
  \tilde{\Psi}_{m}({\bf p} - \frac{{\bf k}}{2})}{m_{N} \left[  (q + k
  - \frac{l}{2})^{2} - M_{V}^{2} + i \epsilon \right]} \\ & + (p
  \leftrightarrow n). \label{eq:double7}
\end{split}
\end{equation}
In Eq.~(\ref{eq:double7}), we have given the amplitude a superscript,
$1$, to distinguish it from the Born term.

We will summarize this section by cleaning up our notation and by
writing out the  correct expressions for the kinematic variables in
terms of the integration variables, taking into account the variable
transformations that were needed to get Eqs.~(\ref{eq:born6a})
and~(\ref{eq:double7}).  We explicitly expand each expression to
linear order in nucleon momentum in the lab frame.   Furthermore, we
assume that nucleon 3-momentum and the exchanged 3-momentum are both
small and of the same order of magnitude relative to all masses
involved.  Subscripts $a$ denote Born amplitude quantities while
subscripts $b$ denote double scattering quantities.  The variables in
each expression are established in the particular  diagram under
consideration.  First, we have,
\begin{equation}
\begin{split}
  \hat{s}_{a} =& \left( \left( P - k  \right) + q\right)^{2} \\ &=
  m_{N}^{2} + 2E_{\gamma}m_{N} + 2E_{\gamma} k_{z} +
  {\mathcal{O}}({\bf k}^{2}) .  \label{eq:kin1}
\end{split}
\end{equation}
Recalling the variable transformations we made in the double
scattering term and noting that ${\bf p}$, ${\bf k}$ and ${\bf l}$ are
all of the same order of magnitude, we have,
\begin{equation}
\begin{split}
  \hat{s}_{b} =& \left( q + P - p_{s} \right)^{2} \\ &= m_{N}^{2} +
  2E_{\gamma}m_{N} + 2E_{\gamma}  \left(p_{z} - \frac{k_{z}}{2}
  \right) + {\mathcal{O}}({\bf p}^2).
\end{split}
\end{equation}
Note that there is only dependence upon $k_{z}$ and that ${\bf k}$
contributions come into play only at higher order in nucleon momentum.
For the rescattering amplitude, we get,
\begin{equation}
\begin{split}
  \bar{s}_{b} =& \left( k_{V} + p_{s} \right)^{2} \\ &= M_{V}^{2} +
  m_{N}^{2} + 2E_{V}m_{N}  -  2E_{\gamma} \left(p_{z} -
  \frac{k_{z}}{2} \right) + {\mathcal{O}}({\bf p}^2).
\end{split}
\end{equation}
This last value is obtained after the pole in $k_{z}$ is taken, giving
the intermediate state an invariant mass of $k_{V}^{2} = M_{V}^{2}$.
The values of $t$ to be used in each of these cases is,
\begin{eqnarray}
  \hat{t}_{a} &=& t \\ \hat{t}_{b} &=& \left( \frac{l}{2} -k
  \right)^{2} =  \left( \frac{l_{0}}{2} \right)^{2} + \frac{l_{z}
  k_{z}}{2} - \left(\frac{l_{\perp}}{2} - k_{\perp} \right)^{2} +
  {\mathcal{O}}({\bf k}^{2}) \label{tkin} \\ \bar{t}_{b} &=&
  \left(\frac{l}{2} +k \right)^{2}  = \left( \frac{l_{0}}{2}
  \right)^{2} - \frac{l_{z}k_{z}}{2} -  \left(\frac{l_{\perp}}{2} +
  k_{\perp} \right)^{2} + {\mathcal{O}}({\bf k}^{2}). \label{kin2}
\end{eqnarray}
In the usual VMD-Glauber theory expression for the double scattering
term, one keeps only the perpendicular components of $\hat{t}$ and
$\bar{t}$.  The terms proportional to $k_{z}$ are small and since they
come with opposite sign, they  tend to cancel if the $t$-dependence of
the basic amplitude is nearly exponential.  The terms with $l_{0}^2$
are proportional to $t^{2}/M_{D}^{2}$.  Thus, we continue to  neglect
both of the first two terms in Eqs. (\ref{tkin}) and (\ref{kin2}).
Finally,
we stress that $(P - k)^{2} = (M_{D} - m_{N})^{2} + {\mathcal{O}}({\bf k}^{2})$ so that
the struck nucleon may be treated kinematically as being on shell up to terms 
quadratic in the nucleon momentum.

By using the kinematic expressions of Eqs.~\ref{eq:kin1}
through~\ref{kin2} in Eqs.~\ref{eq:born6a} and~\ref{eq:double7}, we
may ensure that the factors multiplying the deuteron wavefunction in
each of the integrals is correct to linear order in nucleon 3-momentum
(or exchanged 3-momentum).

\section{Numerical Estimates and the Relationship with VMD-Glauber Theory}
\label{sec:analysis}

\subsection{Differential Cross Section}

Now that we have calculated the Born and Double scattering amplitudes,
let us set up notation that allows us to express the total
differential cross section in terms of the basic amplitudes for
$\gamma N$ and $V N$ scattering.  We do not discuss any physics in this
section, but simply formulate our notation to allow for convenient
comparisons between the present approach and the standard Glauber-VMD approach.   

For any exclusive two body reaction
involving incoming particles of mass $m_1$ and $m_2$ and center of
mass energy squared, s, the differential  cross section may be
represented as follows:
\begin{equation}
  \frac{ d\sigma^{m,m'}}{dt} =  \frac{1}{16 \pi
  \Phi(s,m_1,m_2)}|F_{m,m^\prime}(s,t)|^2, \label{crossection1}
\end{equation}
where,
\begin{equation}
  \Phi(s,m_1,m_2) \equiv  ( (s - m_{1}^{2})^{2} + m_{2}^{4} - 2 s
  m_{2}^{2} - 2m_{1}^{2}m_{2}^{2}). \label{eq:born3}
\end{equation}
In particular, the differential cross section for the reaction in
Eq.~(\ref{reaction}) is,
\begin{equation}
  \frac{ d\sigma^{m,m'}}{dt} =  \frac{1}{16 \pi
  \Phi(s,0,m_N)}|F^0_{m,m^\prime}(s,t) +
  F^1_{m,m^\prime}(s,t)|^2. \label{crossection2}
\end{equation}
It follows from Eqs.~(\ref{eq:born1}) and (\ref{eq:double7}) that the
numerical calculations  of Eq.~(\ref{crossection2}) will require as
input the amplitudes for for both the $\gamma N \rightarrow V N$ and
the $V N \rightarrow V N$ interactions.

To proceed, we construct a parameterization of the photo-production
differential cross section in a form that will provide a smooth
transition to the VMD-Glauber regime by writing,
\begin{equation}
  \frac{d \hat{\sigma}}{d t}^{\gamma N \rightarrow V
  N}(\hat{s},\hat{t}) = \frac{\hat{n}_{0}^{2}}{16 \pi} \left(
  \frac{\hat{s}}{\hat{s}_{0}} \right)^{2(\hat{\alpha}(\hat{t}) - 1)}
  \hat{f}^{2}(\hat{t}) \hat{g}^{2}(\hat{s},\hat{t}) ,
  \label{eq:formfactor1}
\end{equation}
for the basic $\gamma N \rightarrow V N$ interaction.  In the high
energy photon limit, the function $\hat{f}(\hat{t})$  reduces, by construction, to the
usual exponential dependence, $e^{\hat{B} \hat{t}/2}$, with the
constant $\hat{B}$ that is typically  used to parameterize
experimental data as in, for example, Ref.~\cite{Bauer:1977iq}.  The
Regge trajectory is $\hat{\alpha} (t) = \hat{\alpha}^{\prime} t +
\hat{\alpha}_{0}$.  The factor of
$(\frac{s}{s_{0}})^{\hat{\alpha}(\hat{t}) - 1}$ is the Regge
parameterization obtained in the VMD-Glauber regime and
$\hat{g}(\hat{s},\hat{t})$ is a function which adjusts for other $s$
and $t$ dependence that may appear in the intermediate energy regime, but such
that $\hat{g}(\hat{s},0)(\frac{s}{s_{0}})^{\hat{\alpha}(0) - 1}$
reduces to 1 in the high energy photon limit.   By substituting
Eq.~(\ref{eq:formfactor1})  into Eq.~(\ref{crossection1}), we obtain,
\begin{equation}
\begin{split}
  \hat{F}^{\gamma N \rightarrow V N} (\hat{s},\hat{t}) = & \hat{n}_{0}
  (\hat{s} - m_N^2) \left( \frac{\hat{s}}{\hat{s}_{0}}
  \right)^{\hat{\alpha}(\hat{t}) - 1} \hat{f}(t) \hat{g} (\hat{s},\hat{t}) (i +
  \hat{\eta}). \label{eq:formfactor2}
  \end{split}
  \end{equation}
The overall normalization is labeled $\hat{n}_{0}$ and is not necessarily 
related to a total cross section.
The variable, $\hat{\eta}$, is a possible real contribution to the
amplitude.  Because $P_V \gtrsim 1$~GeV for the kinematic regime under consideration 
(see Sec.~\ref{sec:born}), the parameterization we use
for the $V N \rightarrow V N$ simply takes a nearly diffractive form,
\begin{equation}
\begin{split}
  \bar{F}^{V N \rightarrow V N} (\bar{s},t) = & \sigma_{V N}(\bar{s})
  (i + \bar{\eta}) \sqrt{\Phi(\bar{s},m_N,M_V)}
  \bar{f}(\bar{s},\bar{t}). \label{eq:formfactor2b}
  \end{split}
\end{equation}
The function, $\bar{f}(\bar{s},\bar{t})$ reduces by construction to a Regge
parameterization,
$(\frac{\bar{s}}{\bar{s}_{0}})^{\bar{\alpha}(\bar{t}) -
\bar{\alpha}(0)} e^{\bar{B} \bar{t}/2}$ in the VMD regime.  By
applying the optical theorem to Eq.~(\ref{eq:formfactor2b}), we see
that $\sigma_{V N}(\bar{s})$ is, indeed, the total  $V N$ cross
section.  The variable, $\bar{\eta}$ is a possible real part of the
amplitude.

Our peculiar choice of notation is made so that we may smoothly
recover the usual Regge parameterizations  when we consider the
VMD-Glauber approximation.  Indeed, applying the VMD hypothesis in the
appropriate kinematical regime allows us to assume that
$\hat{F}(\hat{s},\hat{t}) \propto \bar{F}(\bar{s},\bar{t})$.  Thus,
applying the optical theorem would allow one to deduce the $V N \rightarrow
V N$ amplitude.  With the standard high energy approximations, we have,
\begin{equation}
\begin{split}
  \hat{F}(\hat{s},\hat{t}) & \stackrel{E_{\gamma} >> M_{V}}{\longrightarrow}  \hat{s}
  \hat{n}_{0} (i + \hat{\eta}) \hat{s}^{\hat{\alpha}^{\prime} \hat{t}}
  e^{\hat{B} \hat{t}/2}  \\ \bar{F}(\bar{s},\bar{t})
  &  \stackrel{E_{\gamma} >> M_{V}}{\longrightarrow}  \bar{s} \sigma_{V N} (i + \bar{\eta})
  \bar{s}^{\bar{\alpha}^{\prime} \bar{t}} e^{\bar{B}
  \bar{t}/2}. \label{amp1b}
\end{split}
\end{equation}

Here, we have put $\hat{s}_{0} = 1$~GeV for convenience 
as is often done in parameterizations.
In this way, we show how our parameterizations reduce smoothly to the
expressions obtained within Regge theory and the VMD hypothesis.

One may fit all of the functions that define the expression  for
$\hat{F}(\hat{s},t)$ directly to data for $\frac{d \sigma^{\gamma N
\rightarrow V N}}{d t}$.   The function, $\hat{g}(\hat{s},\hat{t})$
has been introduced to account for peaks in the energy dependence or
other irregular energy dependence at intermediate energies.   Without the
VMD hypothesis,  we can assume no relationship between
$\hat{F}(\hat{s},\hat{t})$ and $\bar{F}(\bar{s},\bar{t})$.  At intermediate 
energies, therefore, $\bar{F}(\bar{s},\bar{t})$ must be obtained from
a theoretical model or by other experimental means.  Conversely, one can use data for the reaction in
Eq.~(\ref{reaction}) to extract $\bar{F}(\bar{s},\bar{t})$.
\subsection{Corrections to Factorizability and an Effective Form Factor}
\label{sec:corrections}

We now define an  effective form factor,
\begin{equation}
\begin{split}
  S^{m,m^{\prime}}_{eff} \left( E_{\gamma},\frac{{\bf l}}{2}\right)
  \equiv  \int  \frac{d^{3} {\bf k}(\hat{s}_{a} - m_{N}^{2})
  }{2E_{\gamma}m_{N}} \left( \frac{\hat{s}_{a}}{ 2 E_{\gamma} m_{N} }
  \right)^{\alpha(t) - 1} \hat{g}(\hat{s},t)
  \tilde{\Psi}_{m^{\prime}}^{\dagger} \left( {\bf k} - \frac{{\bf
  l}}{2} \right)  \tilde{\Psi}_{m} \left( {\bf k}  \right),
  \label{eq:formfactor3}
\end{split} 
\end{equation} 
for the Born term, and an effective basic amplitude,
\begin{equation}
  \hat{F}^{0}_{eff} (E_{\gamma},t) \equiv  2E_{\gamma}m_{N}
  \hat{n}_{0} \left( i + \hat{\eta} \right)  \left(
  \frac{2E_{\gamma}m_{N}}{s_{0}} \right)^{\alpha(t) - 1}
  f(t). \label{eq:formfactor4}
\end{equation}
If we substitute Eq.~(\ref{eq:formfactor2}) into
Eq.~(\ref{eq:born6a}), then the Born amplitude for production from the
deuteron is,
\begin{equation}
  F_{m,m^{\prime}}^{0}(E_{\gamma},l) = 2
  \hat{F}^{0}_{eff}(E_{\gamma},l) S^{m,m^{\prime}}_{eff,a} \left(
  E_{\gamma},l/2 \right)  + (n \leftrightarrow
  p). \label{eq:formfactor4b}
\end{equation}
The definition in Eq.~(\ref{eq:formfactor4}) takes the form of a
general diffractive  parameterization obtained when one makes the VMD
hypothesis.  However, Eq.~(\ref{eq:formfactor4b}) is exactly correct without
any approximations.  We have recovered the usual structure of the Born
expression - the product of a diffractive basic amplitude with  a form
factor.  The new feature in Eq.~(\ref{eq:formfactor4b})  is that our
effective form factor depends on the energy of the photon.   The
definitions that we made in Eqs.~(\ref{eq:formfactor3})
and~(\ref{eq:formfactor4}) ensure that the effective form factor and
the effective diffractive amplitude reduce to the usual
non-relativistic form factor and the true diffractive  basic amplitude
in the limit that $E_{\gamma} >> M_{V}$:
\begin{equation}
\begin{split}
  S^{m,m^{\prime}}_{eff} \left( E_{\gamma},{\bf l}/2 \right)
  \stackrel{E_{\gamma} >> M_{V}}{\longrightarrow}  S^{m,m^{\prime}}
  \left( {\bf l}/2 \right), \\ \hat{F}^{0}_{eff} (E_{\gamma},l)
  \stackrel{E_{\gamma} >> M_{V}}{\longrightarrow} \hat{F}^{V N
  \rightarrow V N} (\hat{s},t).
\end{split}
\end{equation}
By following the usual methods of VMD-Glauber theory, one
will extract the \emph{effective} amplitude from the $\gamma D \rightarrow V D$ 
cross section rather than
the true  amplitude.  If, in the region of very small $-t$ where the Born 
cross section dominates, the amplitude for the $\gamma N \rightarrow V N$ 
scattering is inferred from data using the usual VMD-Glauber theory, then 
Eq.~\ref{eq:formfactor4b} can be used to obtain a corrected amplitude that accounts for
non-factorizability.
  
\subsection{Corrections to Factorizability in Double Scattering}
\label{sec:doublecorr}

The double scattering term is more complicated due to the fact that,
in Eq.~(\ref{eq:double7}),  the energy dependence cannot easily be
factorized out of the integrand.  We  may rewrite
Eq.~(\ref{eq:double7}) using Eqs.~(\ref{eq:formfactor2})
and~(\ref{eq:formfactor2b}) as,
\begin{equation}
\begin{split}
F^{1}_{m,m^{\prime}} (E_{\gamma},t) = - \int d k_{z} \int d^{2} {\bf
  k_{\perp}} \int \frac{d^{3} {\bf p}}{(2 \pi)^3}  \frac{
  \hat{f}(\hat{t}_{b}) \bar{f}(\bar{s}_{b},\bar{t}_{b})
  \tilde{\Psi}_{m^{\prime}}^{\dagger}({\bf p} + \frac{{\bf k}}{2})
  \tilde{\Psi}_{m} ({\bf p} - \frac{{\bf k}}{2}) }{m_{N}\left[ (q + k
  - \frac{l}{2})^{2}-M_{V}^{2} + i \epsilon \right]} \\  \times \left[
  \left( \frac{\hat{s}}{s_{0}} \right)^{\hat{\alpha}(\hat{t}) - 1}
  \sqrt{\Phi(\hat{s},m_N,0)\Phi(\bar{s},m_N,M_V)} \right] \\  \times
  \hat{g}(\hat{s},\hat{t})  \hat{n}_{0} \sigma_{VN}(\bar{s})  \left( i
  + \hat{\eta} \right)  \left( i + \bar{\eta} \right).
  \label{eq:formfactor5}
\end{split}
\end{equation}
The nonfactorizability of Eq.~(\ref{eq:formfactor5}) near threshold
comes from the fact that the basic amplitudes and the factors in
braces have non-trivial dependence upon the integration variables.
We determine that there is no simple
reformulation of  the integral in Eq.~(\ref{eq:formfactor5}) which
consistently accounts for corrections linear in momentum.   Therefore,
we conclude that a direct numerical evaluation is necessary.  Note that, though
we have set up the integral for a specific parameterization, the
analysis applies to any smooth,  slowly varying energy dependent basic
amplitude.   The $k_{z}$ integral is determined by expanding the
denominator in Eq.~(\ref{eq:formfactor5}),

\begin{eqnarray}
\left( q + k - \frac{l}{2}  \right)^{2}  - M_{V}^{2} + i \epsilon
  \approx & 2E_{\gamma} \left[-k_{z} + \frac{l_{z}}{2} -
  \frac{M_{V}^{2}}{2 E_{\gamma}} + (k - \frac{l}{2})_{0} + i \epsilon
  \right] \nonumber \\ & = 2E_{\gamma} \left[ -k_{z} - \Delta + i
  \epsilon \right]. \label{eq:double8}
\end{eqnarray}
The second line fixes the definition of $\Delta$.  Notice  that by
ignoring the term, $(k - \frac{l}{2})^{2}/2 E_{\gamma}$, we have
ignored the possibility of contributions from intermediate mesons which  are far
off shell and which correspond to nucleon 3-momenta that are strongly
suppressed by the deuteron wavefunction.  Furthermore, note that the
pole value of $k_{z}$ in this approximation only depends  on the
external variables and is independent of the transverse motion of the
nucleons.  The resulting double scattering amplitude is then,
\begin{equation}
\begin{split}
  F^{1}_{m,m^{\prime}}(E_{\gamma},t) = \int d^{2} {\bf k_{\perp}} \int
  \frac{d^{3} {\bf p}}{(2 \pi)^{2}} \int_{-\infty}^{\infty} \frac{d
  k_{z}}{(2 \pi)}  \frac{ \hat{f}(\hat{t}_{b})
  \bar{f}(\bar{s}_{b},\bar{t}_{b})
  \tilde{\Psi}_{m^{\prime}}^{\dagger}({\bf p}  + \frac{{\bf k}}{2})
  \tilde{\Psi}_{m} ({\bf p} - \frac{{\bf k}}{2}) }{2 E_{\gamma} m_{N}
  \left[ k_{z} + \Delta - i \epsilon \right]} \\ \times \left(
  \frac{\hat{s}}{s_{0}} \right)^{\hat{\alpha}(\hat{t}) - 1}
  \sqrt{\Phi(\hat{s},m_N,0)\Phi(\bar{s},m_N,M_V)}
  \hat{g}(\hat{s},\hat{t})   \hat{n}_{0} \sigma_{V N}(\bar{s}_{b})
  \left( i + \hat{\eta} \right)  \left( i + \bar{\eta} \right) \\
  \equiv \int d^{2} {\bf k_{\perp}} \int \frac{d^{3} {\bf p}}{(2
  \pi)^{2}} \int_{-\infty}^{\infty} \frac{d k_{z}}{(2 \pi)}
  \frac{I_{m,m^{\prime}}({\bf k_{\perp}},k_{z},{\bf p},s)}{2
  E_{\gamma} m_{N}  \left[ k_{z} + \Delta - i \epsilon \right]}
\label{eq:formfactor6}.
\end{split}
\end{equation}
We have gathered all factors apart from the energy denominators in the
integrand into a function, $I_{m,m^{\prime}}({\bf
k_{\perp}},k_{z},{\bf p},s)$.   Assuming identical protons and
neutrons, we get an identical term for the case where the roles of the
neutron and proton are inverted.  A convenient way to reorganize this formula so 
that it more closely resembles the non-relativistic quantum mechanical theory
is to write the integrand in terms of its Fourier components in the following mixed representation:
\begin{equation}
  I_{m,m^{\prime}}(k_{\perp},k_{z},{\bf p},s) = \frac{1}{\sqrt{2 \pi}}
  \int_{-\infty}^{\infty} dz \tilde{I}_{m,m^{\prime}}({\bf
  k_{\perp}},z,{\bf p},E_{\gamma}) e^{-i k_{z}
  z}. \label{eq:formfactor6b}
\end{equation}
The vector meson propagator may be rewritten using the identity,
\begin{equation}
  \frac{1}{p - i \epsilon} = \int_{-\infty}^{\infty} dz \Theta (-z)
  e^{i(p - i\epsilon)z}.
\end{equation}
Summing the two terms for the 
neutron and the proton
and using the fact that $\Theta(z) + \Theta(-z) = 1$ yields,
\begin{equation}
\begin{split}
F^{1}_{m,m^{\prime}}(E_{\gamma},t) =&  i \int \frac{d^{3} {\bf p}
  d^{2}  {\bf k_{\perp}} }{2E_{\gamma}m_{N} (2 \pi)^{2} }
  I_{m,m^{\prime}} ({\bf k_{\perp}},-\Delta,{\bf p},E_{\gamma})  \\ &-
  \frac{1}{\sqrt{2 \pi}} \int  \frac{d^{3} {\bf p} d^{2}  {\bf
  k_{\perp}} }{2E_{\gamma} m_{N} (2 \pi)^{2} } \int_{-\infty}^{\infty}
  dz \tilde{I}_{m,m^{\prime}} ({\bf k_{\perp}},z,{\bf p},s)
  \sin(-\Delta z) \Theta(z). \label{eq:formfactor6d}
\end{split}
\end{equation}
In the VMD-Glauber approximation, $\Delta \rightarrow 0$ and
$I_{m,m^{\prime}}({\bf k_{\perp}},0,{\bf p},s)$ is the usual energy
independent density matrix.  Hence, the first term in
Eq.~(\ref{eq:formfactor6d}) reduces to the traditional Glauber
expression for double scattering and the second term vanishes in the limit where
the usual VMD-Glauber assumptions are applicable.  The
second term is a correction, discussed in Ref.~\cite{VM2} which arises
from the non-zero phase shift in the vector meson wave function
induced by longitudinal  momentum transfer.  In the phase shift term,
the factor of $\sin(-\Delta z)$ is itself a correction of order
$k_{z}$, so we neglect Fermi motion and energy dependent corrections
to $I_{m,m^{\prime}}({\bf k_{\perp}},0,{\bf p},s)$ in  the phase shift
term.

For a real photon, the double scattering term picks out the
relative longitudinal nucleon momentum,
\begin{equation}
  \Delta = \frac{l_{-}}{2} + \frac{M_{V}^{2}}{2 E_{\gamma}}.
                           \label{eq:formfactor7}
\end{equation}
Furthermore, $l_{-} = -(M_{V}^{2} - t)/2 E_{\gamma}$, so
\begin{equation}
  \Delta = \frac{t + M_{V}^{2}}{4 E_{\gamma}}, \label{eq:formfactor8}
\end{equation}
and we see that $\Delta$ is indeed negligible at large center of mass
energies and small $t$.  Corrections to the double scattering term, at
linear order in momentum, arise from performing the integral in the
first term of Eq.~(\ref{eq:formfactor6d}) numerically, and by retaining  the phase
shift term.  We end this section by noting that the breakdown in factorization comes
simultaneously from the fact that longitudinal momentum transfer is non-negligible, and the fact that
the longitudinal momentum of the bound nucleons is non-negligible; the contribution to the basic amplitudes from 
the longitudinal component of the bound nucleon momentum at linear order would vanish by 
symmetry in all of the integrals if the longitudinal momentum transfer were neglected in the wavefunctions.

\section{Sample Calculations}
\label{section4}
\subsection{Cross Section Calculation}

It is usually the case that one calculates the charge and quadrapole
form factors in the coordinate space formulation of the form factor.
This method reduces the formulae to an extremely simple form and
allows one to deal simply and directly with polarizations.  For the
purpose of modifying the basic amplitude, however, so that it has
nucleon momentum dependence, we must maintain the momentum space
formulation that results from a direct evaluation of the effective
Feynman diagrams in Figs.~(\ref{fig:figure0},\ref{fig:figure2}).  Carrying 
this procedure out was the topic of the previous two sections.  
The calculation is straightforward, but becomes numerically cumbersome, and the longitudinal momentum 
exchanged leads to a breakdown of the orthogonality relations 
for spherical harmonics that usually lead to a very simple coordinate space formulation. 
However, dealing with the deuteron polarizations can still be simplified if one chooses the axis of
quantization along the direction of momentum 
exchange~\cite{Franco:1969qj,Franco:1965wi}.  An overview of the non-relativistic deuteron
wave function with polarizations are described in Appendix~\ref{sec:app3}. 

In this section we provide some sample calculations by using simple models of the 
basic amplitudes.  To this end, we restore the assumption of VMD and we use very 
simple parameterizations of the $s$ and $t$ dependence in the basic amplitudes.  
The purpose for doing this is mainly to
provide estimates of the sensitivity to non-factorizability rather
than because VMD is thought to be appropriate at intermediate
energies.  We have extracted estimates of the parameters for
production of the $\rho^{0}$ and $\phi$ vector mesons  from the basic
nucleon interaction cross section data appearing in
Ref.~\cite{Bauer:1977iq}, and we have made rough estimates of the
parameterization of the $s$ and $t$-dependence of these amplitudes
(see Sec.~\ref{sec:app4} for a description of our parameterizations).
This provides us with a reasonable model to work with, though we stress that 
refinements are ultimately needed.
For all of our calculations we use the non-relativistic wavefunction obtained from the
Paris N-N potential~\cite{Lacombe:1980dr}.

\begin{figure}
\centering
\rotatebox{270}{{\includegraphics[scale=0.50]{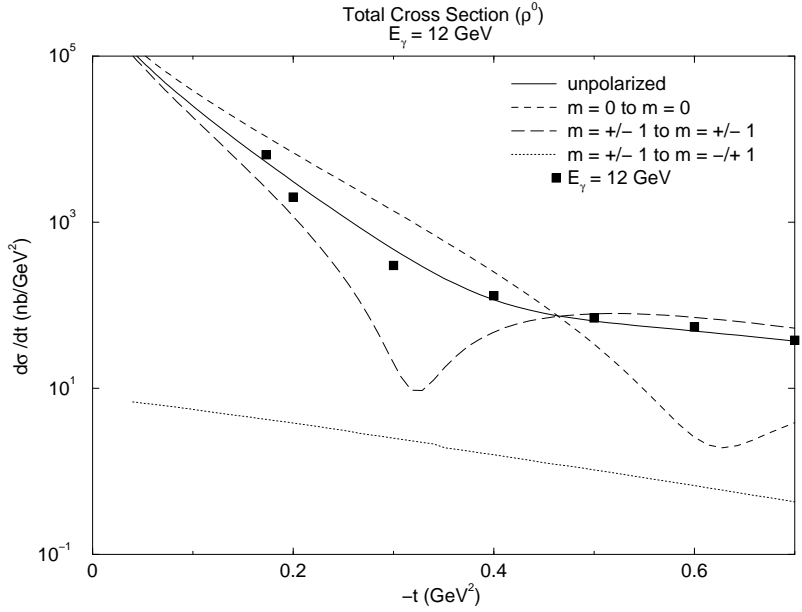}}}
\caption[*]{\label{fig:rho2}The unpolarized differential cross section 
for coherent $\rho^{0}$-meson production compared with the total cross section
for different polarizations.  The calculation is done with the large
photon energy $E_{\gamma} = 12$~GeV, and the data for $E_{\gamma} =
12$~GeV is taken from Ref.~\cite{Bauer:1977iq}.}
\end{figure}

We are mainly interested in the $\phi$-production cross section which
is dominated by natural parity exchange, even at energies close to
threshold, due to the OZI rule.  However, to demonstrate the
consistency of our approach with traditional methods, we consider
first the case of the photo-production of $\rho^{0}$-mesons which has
been well understood for some time.    The basic amplitude for
$\rho^{0}$-production is dominated by soft Pomeron exchange at large
energies, so that it is constant at high energies, but undergoes  a
relatively steep rise at energies near threshold due to meson
exchanges.  The parameterization we use is shown in
appendix~\ref{sec:app4}.  We use a typical exponential slope factor of $7.0$~GeV$^{-2}$ for
the $t$-dependence.   Fig.~(\ref{fig:rho2}) shows the  cross section
for $\rho^{0}$-production at the high energy of $E_{\gamma} =
12.0$~GeV.  For comparison, we show data taken at $12.0$~GeV from
Ref.~\cite{Bauer:1977iq,SLACexp}.  The comparison with data is reasonable, as 
it is with the traditional Glauber approach.

\begin{figure}
\centering
\rotatebox{270}{{\includegraphics[scale=0.60]{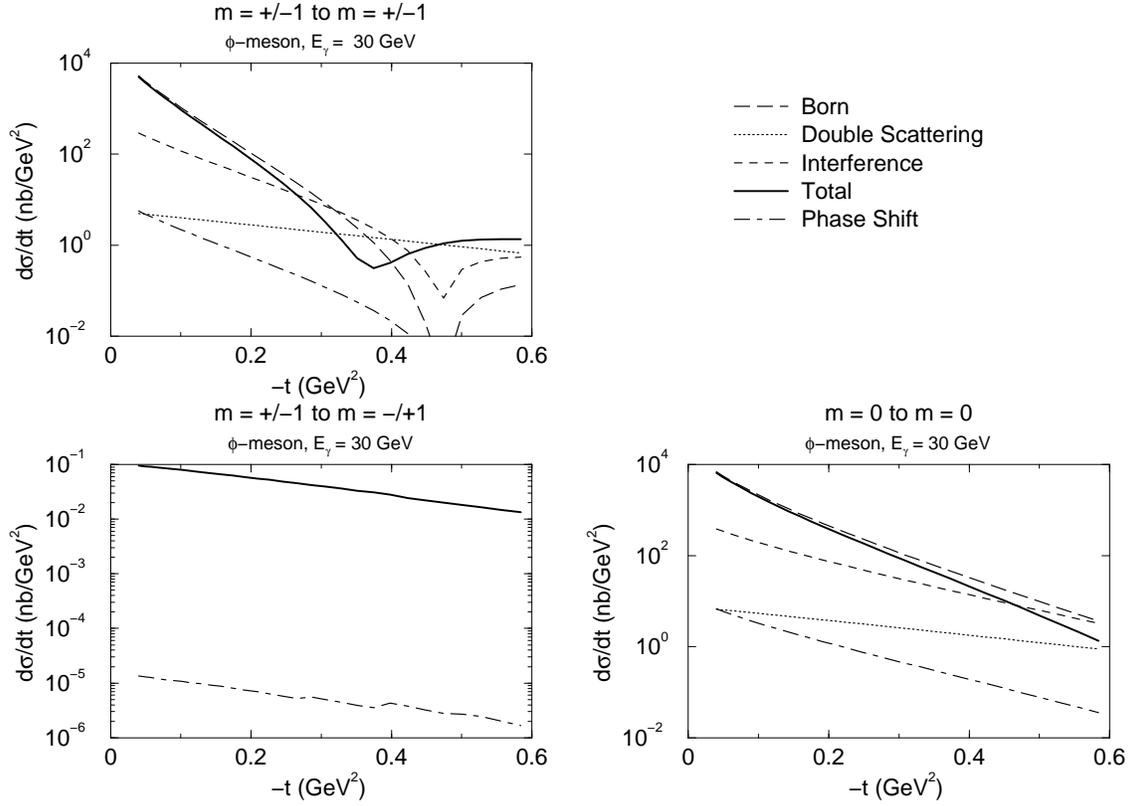}}}
\caption[*]{\label{fig:phi1} 
The long-dashed, dotted, dashed, solid, and dot-dashed lines refer to
the Born, double, interference, total, and phase shift terms
respectively for a photon energy of $E_{\gamma} = 30.0$~GeV.  The interference term is negative but is plotted for illustration on the positive axis.  See text for detailed discussion.}
\end{figure}

\begin{figure}
\centering
\rotatebox{270}{{\includegraphics[scale=0.60]{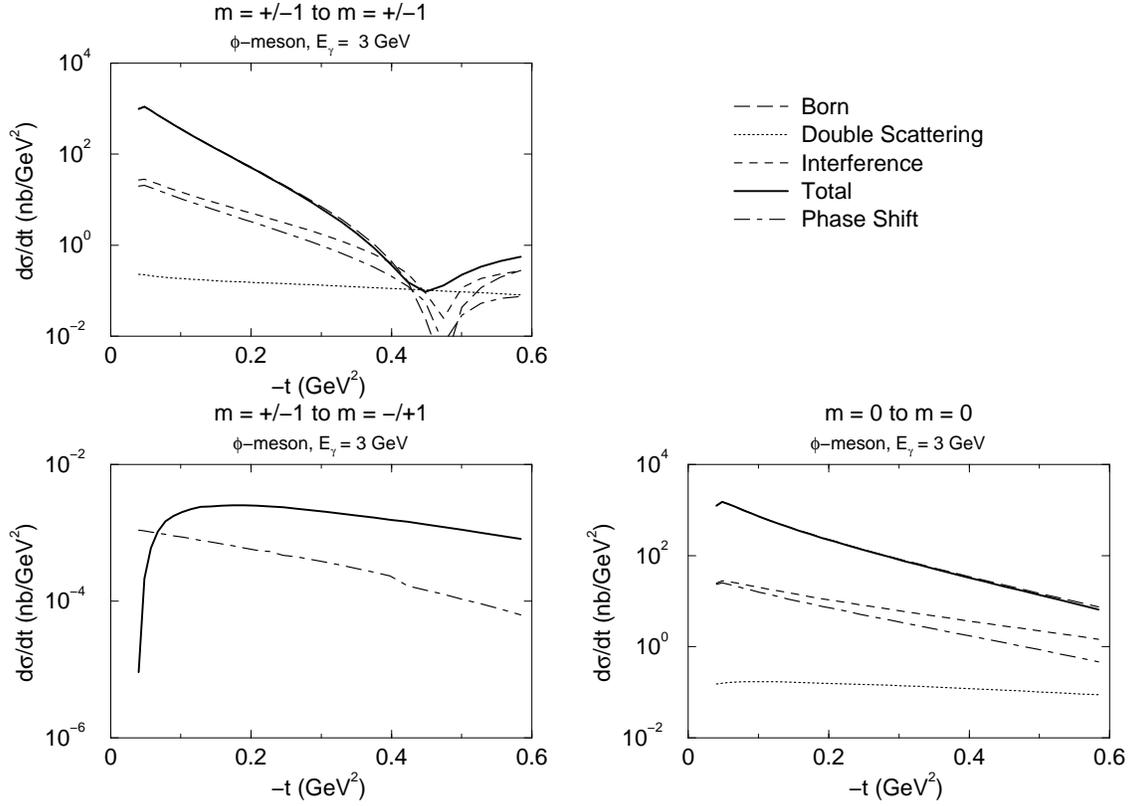}}}
\caption[*]{\label{fig:phi2} 
The long-dashed, dotted, dashed, solid, and dot-dashed lines refer to
the Born, double, interference, total, and phase shift terms
respectively for a photon energy of, $E_{\gamma} = 3.0$~GeV.  Note the different scale on the axis  for
the spin-flip contribution.  The interference term is negative but is plotted for illustration on the positive axis.  See text for detailed discussion.}
\end{figure}

\begin{figure}
\centering
\rotatebox{270}{{\includegraphics[scale=0.60]{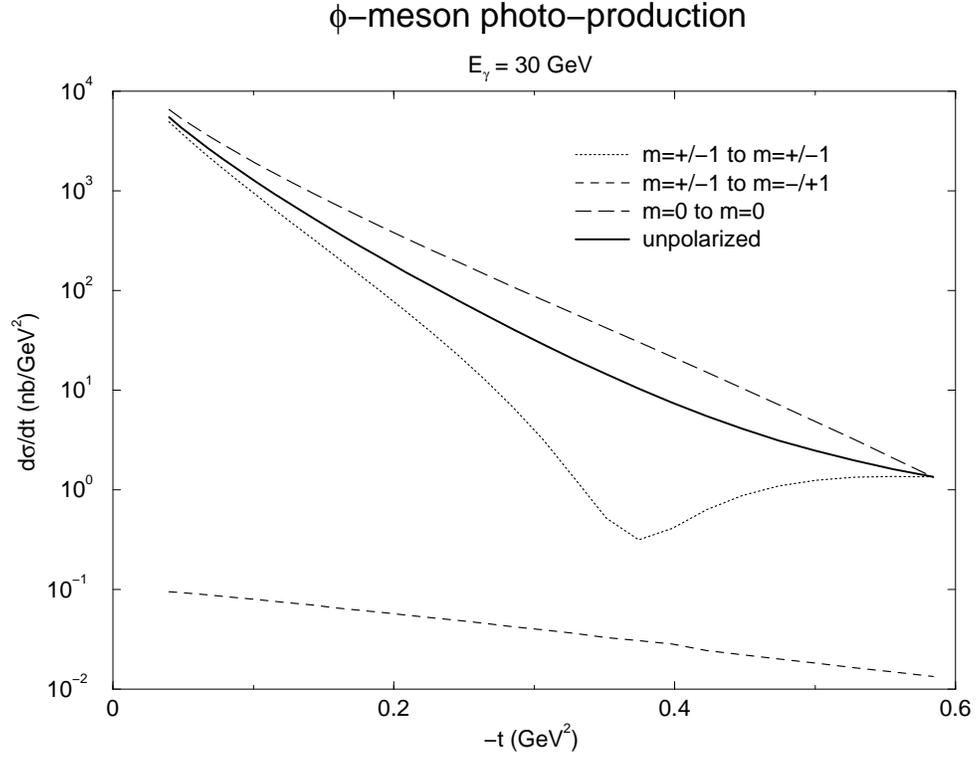}}}
\caption[*]{\label{fig:phi3} The differential cross section for 
$\phi$-meson production for different polarizations for a photon
energy of $E_{\gamma} = 30.0$~GeV.  See text for detailed discussion.}
\end{figure}

\begin{figure}
\centering
\rotatebox{270}{{\includegraphics[scale=0.60]{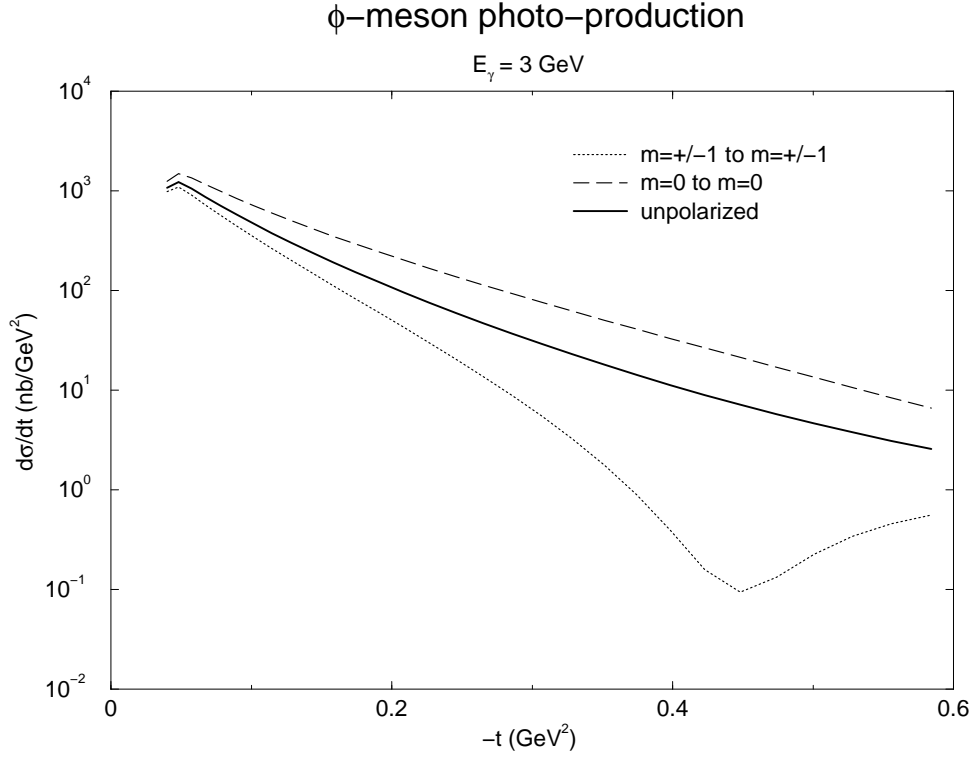}}}
\caption[*]{\label{fig:phi4} The differential cross section for 
$\phi$-meson production for different polarizations For a photon
energy of $E_{\gamma} = 3.0$~GeV.  The deuteron spin flip term is
negligible at these energies.  See text for detailed discussion.}
\end{figure}

Now we consider the more interesting case of $\phi$-meson photo-production.  At high
energy, we use the Regge dependence, $\alpha(t) = .27 t + 1.14$ given
in Ref.\cite{Bauer:1977iq}.  The parameterization that we used is
described further in appendix~\ref{sec:app4}. As noted in
Ref.~\cite{Bauer:1977iq}, the energy dependence of the $\phi$-meson
photo-production cross section is very weak, but the current state of
experimental data is still ambiguous as to how much this energy
dependence continues at lower energies.  However, the large
negative ratio of the real to imaginary  part of the amplitude ($\eta
= -.48$)~\cite{Alvensleben:1971vy} suggests that some mechanism other
than soft Pomeron exchange is significant.  This value of the ratio of
the real to imaginary part of the $\phi$-meson cross section has large
error bars and was calculated neglecting  longitudinal momentum
transfer.  However, it is the only measurement we know of at the
moment so we use it for the  purpose of demonstration.  At lower photon
energies than what we consider here, the energy dependence of the basic cross
section may become highly non-trivial as is suggested by data in
Ref.~\cite{Mibe:2005er}.   The results of the calculation done with
each combination of initial and final deuteron polarizations are shown in 
the separate panels
for a photon energy of $E_{\gamma} = 30.0$~GeV in
Fig.~\ref{fig:phi1} and for a photon energy of $E_{\gamma} = 3.0$~GeV
in Fig.~\ref{fig:phi2}.  
The result is
summarized in Figs.~\ref{fig:phi3} and~\ref{fig:phi4} which show the
differential cross section  for different polarizations along with
the unpolarized cross section for photon energies of $30.0$~GeV and
$3.0$~GeV respectively.

Each of the curves in Figs.~\ref{fig:phi1} and~\ref{fig:phi2} 
separately represents
the contribution to the total cross section from a term in 
the squared amplitude when we apply Eq.~\ref{crossection2}.  The Born and double scattering terms are obtained 
from the square of Eq.~\ref{eq:formfactor4b} and square of the first term of Eq.~\ref{eq:formfactor6d}, respectively.  The phase 
shift term arises from the square of the second term in Eq.~\ref{eq:formfactor6d}.  We call 
it the phase shift term because, in the language of non-relativistic quantum mechanical
wavefunctions, it arises due to a phase difference  between the incoming photon and the produced
vector meson.  The interference term arises from the interference between Eq.~\ref{eq:formfactor4b} and
Eq.~\ref{eq:formfactor6d}.  Note that the interference term is negative, but it is plotted on the positive
axis for demonstration purposes.
Note also that there is no contribution from the
Born term for the $m=+/-1$ to $m=-/+1$ transition, and therefore the total cross section for the spin-flip reaction has none
of the large dips characteristic of the Born cross section. 

An important feature that can be seen in Figs.~\ref{fig:phi1} and~\ref{fig:phi2} is that the double 
scattering term is suppressed in the intermediate energy case relative to the high energy case.
We can see this most clearly by comparing the upper left panel of Fig.~\ref{fig:phi1} with the upper left panel of Fig.~\ref{fig:phi2}, 
It is clear that the double scattering contribution is important in the $E_{\gamma} = 30$~GeV case at moderate
values of $-t$, whereas for the $E_{\gamma} = 3$~GeV case the cross section is dominated by the Born term
all the way up to $-t \approx .4$~GeV$^2$.  In the general case of multiple scattering from complex nuclei, it 
is the rescattering contributions which lead to the usual A-dependence (A is the number of nucleons) of Glauber theory.
The fact that multiple scattering is suppressed in double scattering in the deuteron suggests that our method would
yield a rather different A-dependence from that of usual Glauber theory if it were extended to complex nuclei.  
Extending our approach to complex nuclei will be the subject of future work.  

Another problem begins to emerge at lower photon energies and extremely small $-t$ (at $t \approx t_{min}$):  A
large fraction of the momentum integrals begins to violate
relativistic kinematic constraints.  It is likely that the basic
amplitudes vary extremely rapidly with $s$ and $-t$ in these regions of the integral and
that expanding in nucleon momentum is not valid (at least to linear
order).  In order to make progress, a precise understanding of the dynamics of
off-shell amplitudes based on field theory may be necessary.  Therefore, our approximation
is only valid at $-t$ sufficiently large that the integrand does not contain
significant contributions from kinematically forbidden nucleon configurations.
We have tested the effect of this region in our calculations, and   
in performing our calculation, we find that there is virtually no contribution
from kinematically forbidden regions for any situation that we consider as long as $-\hat{t} + \hat{t}_{min}$ is greater than a few tens of MeVs.
We note that, even at relatively low photon energies, the data is consistent with a smooth exponential $-t$-dependence
(see Ref.~\cite{Mibe:2005er} ) as long as $-t$ is not exactly $-t_{min}$. Note that this
theoretical problem of considering $t$ at exactly $-t_{min}$ exists at high energies as well, but that at high energies $-t_{min}$ is generally too small for it to show up in plots.  
So that we may perform our calculations numerically at all values of $-t$ greater than $-t_{min}$, we
choose to make the basic amplitude vanish in kinematically forbidden configurations (when $-t \leq -\hat{t}_{min})$.  
This results in a small dip just above $-t_{min}$ in our plots.  The small dip is, therefore, unphysical, and should not be regarded as a prediction.  We leave
it in our plots merely to illustrate a general failure of the Glauber theory approach at extremely small $-t$ (see Fig.~\ref{fig:phi4} at $-t \lesssim .06$~GeV$^{2}$).  In 
summary of the above, the 
small dip at extremely small $-t$ denotes a kinematic region in which no known multiple scattering formalism works.  Numerically, 
our calculation is only correct in the region of $-t$ above the dip at small $-t$; that is, when $-\hat{t} + \hat{t}_{min}$ is greater than a few tens of MeVs.  

\clearpage

\clearpage

Next we plot consider the total unpolarized cross sections as a function of
photon energy for a set of fixed values of $-t$.
This allows us to
compare the factorized and unfactorized calculations directly and to
determine at approximately what value of energy the transition to the
VMD-Glauber regime occurs.  
Recall that it is the motion of the nucleons in the deuteron (the Fermi motion) 
that leads to the non-factorizability of the basic amplitudes.  Factorization refers the practice of ignoring the
dependence of nucleon momentum inside the basic amplitudes when integrals over nucleon momentum are performed.
The use of 
non-factorized amplitudes is the essential difference between our approach and 
the usual Glauber approach.
\begin{figure}
\centering \rotatebox{270}{{\includegraphics[scale=0.80]{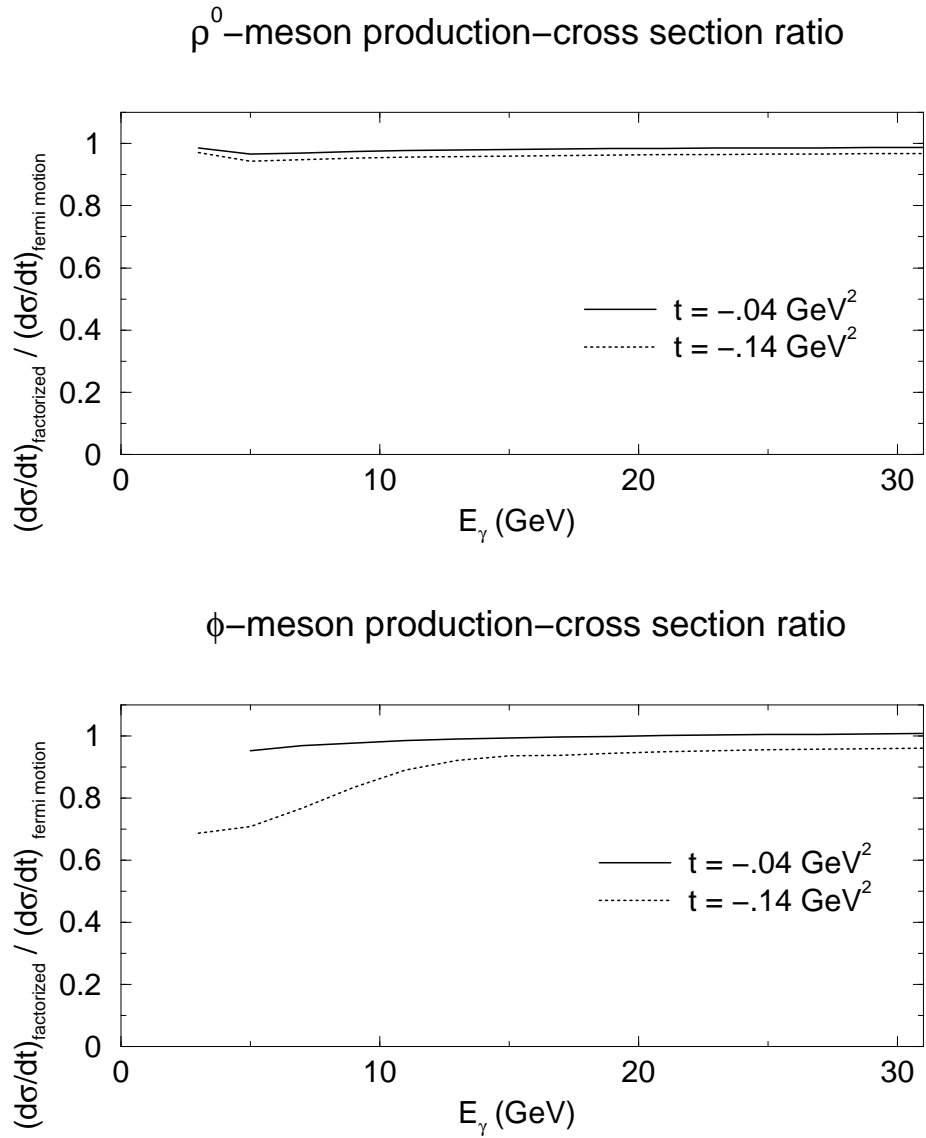}}}
\caption{The energy dependence of the ratio between the differential cross section 
calculated using the usual factorization assumption and the differential cross section
calculated with factorization break-down taken into account.}
\label{fig:figure9}
\end{figure}
The ratio of the cross section with the 
usual factorization assumption to the cross section which accounts for non-factorizability effects (Fermi motion) is shown in Figs.~\ref{fig:figure9}.
The upper panel refers to the case of $\rho^{0}$ production, whereas the lower panel refers to $\phi$-production.  The ratio 
is given for two small values of $-t$: $t=-.04$~GeV$^2$ and $t=-.14$~GeV$^2$.

The upper panel demonstrates that the effect of non-factorizability is small for the case of $\rho^{0}$ for the entire range
of intermediate energies.  This is in sharp contrast to the case of $\phi$ production in the lower panel of Fig.~\ref{fig:figure9}.
Note that we only plot the case of $t=-.04$~GeV down to $E_{\gamma} = 5$~GeV for the $\phi$-meson case.  This is because, for photon energies lower than $5$~GeV, $t=-.04$~GeV becomes
too close to $t_{min}$.  On the other hand, for the curve corresponding to $t=-.14$~GeV, 
there is nearly a 30 percent suppression
of the factorized cross section relative to the unfactorized cross
section at the lowest energy, $E_{\gamma} = 3$~GeV, shown in the lower panel of Fig.~\ref{fig:figure9}.  
We emphasize that this result is for a photon energy ($3$~GeV) that is well into the kinematic
region where the eikonal approximation may be applied (see Sec.~\ref{sec:born}), and that $-t =.14$~GeV$^2$ is certainly 
large enough relative to $-t_{min} = .036$~GeV$^2$ that there are none of the problems discussed earlier related to
nearness to $-t_{min}$.  Therefore, our method of calculation is ideally suited to the kinematics of the dotted curve in Fig.~\ref{fig:figure9}, where 
a significant effect from the break down of the factorizability assumption is already seen.

Note from the general behavior in Fig.~\ref{fig:figure9} that the cross section rises
when the factorization assumption is removed.  This effect is mainly due to the suppression of 
multiple scattering when non-factorizability is taken into account.  To see this, note that At $-t = .14$~GeV$^2$ the main effect of double scattering in the usual Glauber approach 
is to produce a 
large negative cross term that has a canceling effect.  All terms apart from the Born term and the interference term
are negligible at this value of $t$.  (See, for example, the upper left panels of Figs.~\ref{fig:phi1} and~\ref{fig:phi2}.)
Therefore, if multiple scattering is suppressed, as it is in our approach, then the absolute value of the 
cross term becomes smaller, and the Born contribution is no longer suppressed by multiple scattering.  
Thus, the curve representing our approach in Fig.~\ref{fig:figure9} is smaller than what is found in the standard Glauber calculation.
  
At high energies we expect the two methods to agree, and they do within the range of experimental uncertainties of non-relativistic deuteron
form factors.  The fact that the two methods have slight disagreement at high energies is a reflection of the fact that, even at high 
energies, we have not calculated the form factor with exactly the same approximation as in the usual Glauber approach.  In the usual non-relativistic
form factor, any dependence on longitudinal transferred momentum is ignored.  If one takes into account exact kinematics, one finds that there are 
two distinct effects which may cause this assumption to be violated.  It is easiest to see this by writing out the exact expression for the transferred 
longitudinal momentum:
\begin{equation}
l_{z} = -\frac{t}{2 M_{D}} + \frac{M_{V}^{2} - t}{2 E_{\gamma}}.
\end{equation}
From this we see that there are two approximations that are normally made in the Glauber approach that allow 
one to neglect $l_{z}$.  
The first is the ultra-relativistic approximation for the incident vector meson, $E_{\gamma} >> M_{V}$, and 
the second is the non-relativistic approximation for the exchanged 4-momentum, $-t << M_{D}$.  If $t$ is small 
relative to $M_{D}$ then there is still a significant contribution to $l_{z}$ when $M_{V}$ is non-negligible relative to $E_{\gamma}$.
This is the effect that interests us in this paper.  It is safe to use the non-relativistic form factor because the transferred energy is, 
\begin{equation}
l_{0} = -\frac{t}{2 M_{D}},
\end{equation}
which is small at small $-t$.
On the other hand, as long as $t$ is not exactly zero, there will be a component of $l_{z}$ that does not die out with energy.  This effect represents
the error induced by ignoring relativistic recoil.  In the future we plan to generalize the formalism to the case of light-cone wave functions so 
that it may be extended to higher $-t$. 

\section{Directions for Future Work}
\label{sec:future}

\subsection{Extraction of Basic Amplitudes}
\label{sec:basicamps}
\begin{figure}
\centering \rotatebox{270}{{\includegraphics[scale=0.50]{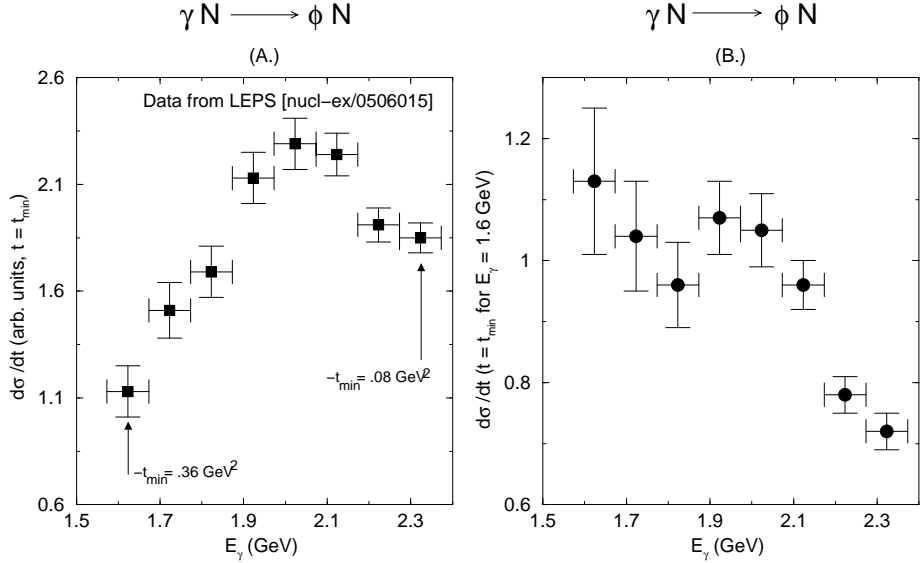}}}
\caption{Plot of recent data from LEPS, taken from Ref.~\cite{Mibe:2005er}.  We indicate the
significant variation of $t_{min}$ with photon energy.  This may have an effect on the over all
energy dependence of the cross section.  (A.) shows how the data was originally presented: at a different value of $t$ for each energy.  
In (B.) we have shifted all of the data points to the same value of $t$ by assuming a constant slope parameter of $4$~GeV$^{-2}$.  Each point
in (B.) corresponds to the differential cross section at the fixed value of $t$ corresponding to to $t_{min}$ for a $1.6$~GeV photon.  Note the different scales on 
the axes in (B.)}
\label{fig:lepsdata}
\end{figure}

We emphasize that the work in this paper is a first step in refinements to the usual techniques 
applied to multiple scattering in vector meson production from the deuteron.  We plan to extend 
these refinements in the future to include, for example, light-cone kinematics in the treatment of the
deuteron wavefunction, and spin-flip effects.
Obtaining precise parameterizations of the $s$ and $t$ dependence is one of several
steps needed for refinements in the calculation.
We note that a peak in the energy dependence has been 
reported in Ref.~\cite{Mibe:2005er} for photo-production of $\phi$-mesons from a proton target at $E_{\gamma} = 2$~GeV and it is this data that we
used in our parameterization (see Appendix~\ref{sec:app4}).  We would
like to point out, however, that the measurements in Ref.~\cite{Mibe:2005er} are for the differential cross section at $t = t_{min}$.  Therefore, since 
the value of $t_{min}$ varies significantly with energy in these near threshold measurements, then the reported measurements give  
the differential cross section at very different values of $t$.  We have indicated this in Fig.~\ref{fig:lepsdata}(A.). In order to infer the energy dependence at a \emph{fixed} value of
$t$, one needs to assume a form for the $t$-dependence.  The actual $t$-dependence at these low energies is not well known, but it is straightforward to
see that even a simple exponential $t$-dependence will have an effect on the shape of the over-all energy dependence of the cross section.  As an example, we 
have plotted in Fig.~\ref{fig:lepsdata}(A.) the data as it was originally presented in  Ref.~\cite{Mibe:2005er} alongside Fig.~\ref{fig:lepsdata}(B.) where the data have been 
shifted to a fixed value of $t$.  We have used an exponential slope parameter of $4$~GeV$^2$ which gives reasonable agreement with the data.  In the original form of the plot, Fig.~\ref{fig:lepsdata} (A.), the
data is shown at a different value of $t$ at each energy.  The highest value of $-t_{min}$ occurs at the lowest energy plotted which is around $1.6$~GeV. In Fig.~\ref{fig:lepsdata}(B.) we have re-plotted the 
energy dependence, but with the value of $t$ for each data point fixed at $-t_{min}$ for $E_{\gamma} = 1.6$~GeV since this is the largest value of $-t$ that is kinematically allowed
for every point on the plot.  Let $t_{min}[1.6]$ represent
the value of $-t_{min}$ at $E_{\gamma} = 1.6$~GeV. Then,
\begin{equation}
\left. \frac{d \sigma}{dt} \right|_{t = t_{min}[1.6]} = \left. \frac{d \sigma}{dt} \right|_{t = t_{min}} e^{4.0 {\mathrm GeV}^{-2} (t_{min}[1.6] -  t_{min})}. 
\end{equation} 
We use this to obtain Fig.~\ref{fig:lepsdata} (B.).
We see that much of the peak-like behavior is
removed.  Without a fuller understanding of 
the $t$-dependence, therefore, it cannot be ruled out that the observed peak arises from purely kinematical effects.
However, the fact that the cross section at fixed $t$ does increase at smaller $E_{\gamma}$ is evidence that OZI-violating meson exchange effects become important at these energies.  

Recently, preliminary data were reported from SPring-8/LEPS~\cite{PANIC}  
which measured the dependence of the $\phi$-meson production cross section at $E_ 
{\gamma}\sim 2$~GeV on the linear polarization of the photon.  
Significant polarization is observed which requires the presence of a  
non-vacuum exchange like $\pi,\eta$ exchange.  Such exchanges lead to  
spin flip in the nucleon vertex.  These contributions for small t are  
strongly suppressed for coherent production off the deuteron (pion  
exchange does not contribute in any case due to the zero isospin
of the deuteron.)  These effects are determined by the deuteron magnetic form  
factor  which is much smaller than the electric form factor. Hence,  
the coherent production of the $\phi$-meson may be used as a spin  
analyzer of the elementary amplitude in the kinematics where double  
scattering is a small correction.
This topic has already been discussed in Ref.~\cite{Titov:2002zy}.
In the spirit of the original Glauber approach, we have neglected spin effects in 
this paper for the sake of simplicity.
Future work will involve generalizations of our method to the
case of spin dependent basic amplitudes.  However, if one fits a combination 
of Pomeron trajectory and Reggeon trajectory to the recent preliminary SPring-8/LEPS data, and then 
extrapolates to $3$~GeV, then it appears that less than $20\%$ of the basic $\gamma D$ cross section
is due to spin-flip, whereas the corrections found in this paper due to non-factorizability are as
large as $30\%$ at $3$~GeV~\cite{PANIC}.

\begin{figure}
\centering \rotatebox{270}{{\includegraphics[scale=0.50]{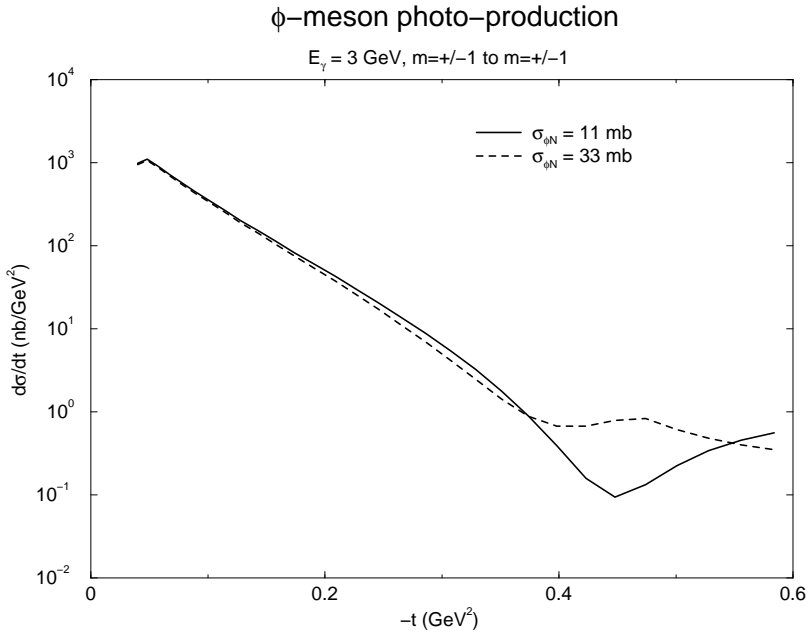}}}
\caption{The energy dependence of the $m=+/-1$ to $m=+/-1$ differential 
cross section for $\phi$-meson photo-production with $E_{\gamma} =
3$~GeV.  The dashed curve shows the result of increasing the typical
basic $\phi N$ cross section by a factor of 3.}
\label{fig:unnamed}
\end{figure}

Before ending our analysis, we mention that, because $V-N$ cross sections
are extracted from the multiple scattering term, quantities sensitive to the 
deuteron polarization would be ideal for testing whether the $V-N$ cross section
is unusually large. 
In order to emphasize this, the
cross section for scattering from a polarized  deuteron, from $m=+/-1$ to
$m=+/-1$, is plotted in Fig.~\ref{fig:unnamed} where the result of
using a typical value for the total $\phi N$ cross section ($11$~mb)
is compared with the case when the $\phi N$ cross section is enhanced
by a factor of three.  (For clarity we have only plotted the sum of all the terms 
from the squared amplitude rather than each term separately.)  
The sharp dip that normally appears, is due to the sharp dip in the 
Born cross section.  However, the double scattering cross section is nearly flat in $-t$.  Therefore, in the summed cross section the double scattering term
dominates in the region of the dip, and may even cause the dip to vanish entirely if it becomes very large.
Figure~\ref{fig:unnamed} shows that,
even with the suppression of the double
scattering term that results from the non-factorizability that we have been discussing, 
the dip in the cross
section is observed to flatten out when the basic $\phi N$ cross
section is abnormally large.  

\subsection{The Problem of Bound State Amplitudes}
\label{sec:off-shell}

We have treated the struck nucleon as being on-shell which is
consistent with the neglect of terms quadratic in nucleon momentum.
However, \emph{immediately} at the threshold for particle production,  the
$\gamma N \rightarrow V N$ amplitude has very unpredictable behavior which may be modified
significantly  when the nucleon is in a bound state.  This is
especially clear when we  realize that for a given photon energy
$-t_{min}$ is different for a deuteron and an on-shell nucleon target.
We cannot predict the effects of the off-shellness of the bound
nucleon without a complete, relativistic understanding of the basic
amplitude.  However, we have made predictions in the region of
kinematics where it is reasonable to assume that the bound state
amplitude is the same as that of the free nucleon amplitude.  If one includes 
 dependence upon the nucleon virtuality in the basic amplitude, then one may 
write the amplitude as $\hat{F}(\hat{s},\hat{t},k_{N}^{2})$.  For $k_{N}^{2} = m_{N}^{2}$, 
the amplitude reduces to the free nucleon amplitude.  As we have stated,
$k_{N}^{2} = m_{N}^{2}$ up to corrections of order ${\bf {k}}_{N}^{2}/m_{N}^{2}$ or
higher whereas $\hat{s}$ has linear order corrections in nucleon momentum.  Thus, if 
$\hat{F}(\hat{s},\hat{t},k_{N}^{2})$ is an analytic function of kinematic variables, then 
there will be linear order corrections in nucleon momentum due to $\hat{s}$ whereas the 
lowest order corrections due to the virtuality of the bound nucleon are only of quadratic 
order in nucleon momentum.  In other words, Fermi motion effects may be important even when 
it is appropriate to neglect the off-shellness of the bound state amplitude.  
Of course, all of this depends on the validity of using ${\bf {k}}_{N}/m_{N}$ as
a small expansion parameter which is only true if the basic amplitude has relatively weak
s-dependence.  This is one reason why we emphasize that we are considering intermediate energies rather than
low energies.
One may also include the deuteron binding energy in the calculation
of the mass of the bound nucleons, but the binding energy arises from the full consideration
of relativistic binding and higher order terms in nucleon momentum, so considering the nucleon
binding energy is not consistent with the neglect of higher order nucleon momentum terms or
the use of a non-relativistic potential for the N-N interaction. 

In this subsection we propose a rough a way to test the validity of the
on-shell amplitude approximation. We do this in the next few
paragraphs by directly comparing the amplitude  when it is evaluated
at the value of $t_{min}$ for the deuteron with the case when it is
evaluated at $t_{min}$ for a free nucleon with $s$ given by the exact
expression for $\hat{s}$,
\begin{eqnarray}
  \hat{s} = 2 E_{\gamma} \left( M_{D}  - \sqrt{m_{N}^{2} + {\bf
  k}^{2}} + k_{z} \right) + \left( M_{D} - \sqrt{m_{N}^{2} +   {\bf
  k}^{2}} \right)^{2} - {\bf k}^{2}. \label{eq:shat}
\end{eqnarray}
The value of the nucleon 3-momentum thus parameterizes the
off-shellness of the bound nucleons.   $t^{\prime}_{min}$ will denote
the lower bound of $-t$ for the free nucleon, whereas $t_{min}$  is
the lower bound of $-t$ for the deuteron.  The struck  nucleon  inside
the deuteron for the unprimed case has $\hat{s}$ given by
Eq.~(\ref{eq:shat}).  We will now consider the case of a free nucleon,
with the same $\hat{s}$ as for the bound nucleon, but with the nucleon
on-shell (i.e. $k^{2}=m_{N}^{2}$) and with a fixed value for $k_{z}$.
So that the free nucleon energy corresponds to the bound nucleon
energy, we will continue to use $M_{D} - \sqrt{m_{N}^{2} + {\bf
k}^{2}}$ for the energy of the struck nucleon.  In short, we are
comparing $t_{min}$ for $\gamma$ scattering off a deuteron at rest with
$t^{\prime}_{min}$ for $\gamma$ scattering off a free nucleon, with
energy corresponding to that of the bound nucleon in both cases.

We expect the rate of variation of the basic amplitude with $t$ to be
very large near $t_{min}$.  If there is a significant contribution to
the integral in Eq.~(\ref{eq:born6a}) from regions near $t_{min}$,
then $t_{min}$ should  nearly equal $t^{\prime}_{min}$ in order to
make the on-shell amplitude a valid approximation to the bound state
amplitude.   We can use the difference between these two values of
$t_{min}$ to  estimate the effect of the the off-shellness on the the
amplitude.

In order to test the effect of the off-shellness of the basic
amplitude, we may consider two extremes.   First, the bound state basic
amplitude could be evaluated at the physical value of $t$ for the
photon-deuteron  process.  That is, we could calculate the amplitude,
$\hat{F}(\hat{s},t)$ at $t$ where $t$ is the physical value of $t$ for
the photon deuteron process.   In this case, since $t_{min}$ is
smaller for the deuteron than $t^{\prime}_{min}$ is for the nucleon,
then we are probably over-estimating the cross-section.  On the other
hand, we could evaluate the  basic amplitude at $F(\hat{s},t -
(t_{min} - t_{min}^{\prime})) = F(\hat{s}, t - \Delta t)$  where
$t^{\prime}_{min}$ is the minimum $t$ for the free, on-shell nucleon.
With this second method for choosing which value of $t$ to use in the
basic amplitude, the basic amplitude behaves like the free, on-shell
nucleon amplitude in the region of $t$ close to $t_{min}$.  Hence,
with this method, we are probably underestimating the value of the
basic amplitude.  In the high energy limit, $\Delta t$ vanishes and
the two amplitudes are equal, and the difference between the two
provides an  estimate of the off-shell effects.  (Note that we must
specify a value for ${\bf k}$ in order to make a comparison.)  Any
amplitude which has a relatively slow and smooth variation with $t$
will yield a small difference between $\hat{F}(\hat{s},t)$ and
$F(\hat{s}, t - \Delta t)$.  We made this comparison for $\phi$-meson
production with the parameterization in appendix~\ref{sec:app4} and we
find  only a few percent deviation.  We conclude that at a few GeV
above threshold it is reasonable to continue using the on-shell
amplitude of the nucleon.  

\section{Summary and Conclusions}
\label{sec:summary}

The main conclusion of this paper is that the effect of factorization
break-down is significant for intermediate photon energies.  The Glauber approach is, 
strictly, only applicable for the case of very high photon energies.  However, there are 
current attempts to apply the factorization assumption of Glauber theory to the $\phi$-meson production 
reaction at energies as low as $1.5$~GeV in both experimental and theoretical research.  
Therefore, in order to salvage the situation in the energy range of a few GeVs above threshold, 
we have outlined
steps one must follow in order to obtain corrections to leading
order in the bound nucleon momentum and transferred momentum.  The main steps are essentially 
those of the original 
diagrammatic formulation of Glauber theory in terms of momentum space integrals
and its extension to vector meson production~\cite{Gribov:1968jf,Kolbig:1968rm,Bertocchi}; we have started with most of the 
original assumptions, but we have removed the assumptions of factorizability, ultra-relativistic 
kinematics or VMD for 
the basic amplitudes, and we have numerically evaluated all integrals directly without any factorization approximations.     
By using a simple model for the basic amplitude (we restore VMD for the simple model) based on a fit to old and recent data, 
we have shown that, away from $t = t_{min}$, ignoring
Fermi motion (and the resulting breakdown of factorizability) can lead 
to a significant error in basic cross
sections extracted from $\gamma D \rightarrow \phi D$ cross section
data (see
Fig.~\ref{fig:figure9}).  This effect will certainly need to be taken into account in
future searches for new production mechanisms at intermediate energies.  
The breakdown in factorizability arises as a consequence of both the non-negligible 
longitudinal momentum exchanged, and the non-negligible Fermi motion.
An important point is that a key source of 
the departure from factorizability is the inadequacy of assuming the nearly 
flat s-dependence predicted by Regge theory in the basic amplitudes.  Therefore, models 
of the basic $\gamma N \rightarrow V N$ amplitude or the $V N \rightarrow 
V N$ amplitude which depart significantly from nearly flat s-dependence  must include 
\emph{at least} the linear order nucleon momentum 
corrections of this paper if they are to 
be used in calculations with a deuteron target.  This correction arises purely from the fact that
the bound nucleons have non-vanishing momentum 
and it must be included regardless of the details of a particular model of the 
basic amplitudes.  For the case of $\phi$-meson production, we find that our  
approach is reasonable 
when we use our particular simple model of the 
basic amplitude and as long as the photon energy is around $3$~GeV or higher and $t$ is not too close to $t_{min}$.  

However, we stress that in a model of the basic $\gamma N \rightarrow V N$ amplitude that predicts much wilder energy dependence at intermediate energies than what we
have assumed, the linear order corrections will not be sufficient, and a complete and precise
understanding of the $N-N$ interaction and the bound state nucleon amplitudes are necessary in order to make 
a correct calculation.  For the $\phi$-meson photo-production cross section ($M_{\phi} \approx 1.02$~GeV), 
the amplitude may vary wildly with energy at $E_{\gamma} = 2$~GeV or lower because of the very close proximity to threshold.  For this reason, and because the
eikonal approximation begins to break down, basic cross sections for photo-production from the nucleon extracted from data for photo-production from 
the deuteron are suspect for 
photon energies less than or equal to $2$~GeV for the $\phi$ production reaction.  

In our sample calculation, we observe that the contribution from double scattering
becomes numerically suppressed relative to the Born approximation as the incident photon energy decreases.  However, the multiple scattering terms are what 
lead to the characteristic A-dependence of the Glauber theory for complex nuclei, $\sigma_{tot} \sim A^{2/3}$. 
This suggests that an extension of our methods to complex nuclei will yield a rather 
different A-dependence for the cross section at intermediate energies from what is predicted at high energies.  Hence, there
will need to be a revision in efforts to extract basic cross sections
from  nuclear data using extrapolations in $A$.  The extension to complex nuclei, however,
requires much more work.  We note, however, that data
given in Ref.~\cite{Ishikawa:2005aw} were interpreted as implying  a very high $\phi N$
total cross section on the basis of a very traditional Glauber approach at energies of only a few GeVs.  Therefore, our next
step will be to determine how the non-factorization effects discussed
in this paper affect a general, incoherent Glauber series.  Furthermore, since it is apparent that spin effects will 
be important, then a generalization with spin-dependent amplitudes will be needed.

We have purposefully over-simplified our analysis here for the
purposes of demonstration.  In particular, we have applied the VMD
hypothesis at energies where it is suspect and we have neglected
fluctuations and $\omega-\phi$ mixing in  the intermediate vector
meson in the double scattering term.  Further analysis will need to
include these effects.  In order to make further numerical progress, we
will need firmer parameterizations of the basic cross sections for
vector meson production from nucleons.  For theoretical
work, it would be useful for the purposes of comparison to have a
widely agreed upon set of parameterizations.  We also need to consider the calculation 
in light-cone coordinates and the effects of spin-flip.  We will pursue these issues in future work.

Finally, we will need a complete understanding of the off-shell
amplitudes if we are to take into account the  higher order momentum
corrections that will be necessary just at the threshold, though we
have argued that for smoothly varying basic cross sections, the
effect of off-shellness in the amplitudes is small relative to the effect of
linear order corrections in nucleon momentum.

\begin{acknowledgments} 
T.C. Rogers thanks Steve Heppelmann and David Landy for useful discussions,
and Isaac Mognet and Nick Conklin for sharing computer facilities.  This work is supported by DOE grants under contract DE-FG02-01ER-41172
and DE-FG02-93ER40771.
\end{acknowledgments}

\appendix

\section{Parameterizations}
\label{sec:app4}

Here we describe the fits of the basic 
cross sections that we used for our sample calculations.  The object
here is not necessarily to produce very accurate  parameterizations, but
rather to devise parameterizations that demonstrate the effects of
Glauber factorization while being consistent with recent and
established experimental results.

\begin{figure}
\centering \includegraphics [scale=1.3]{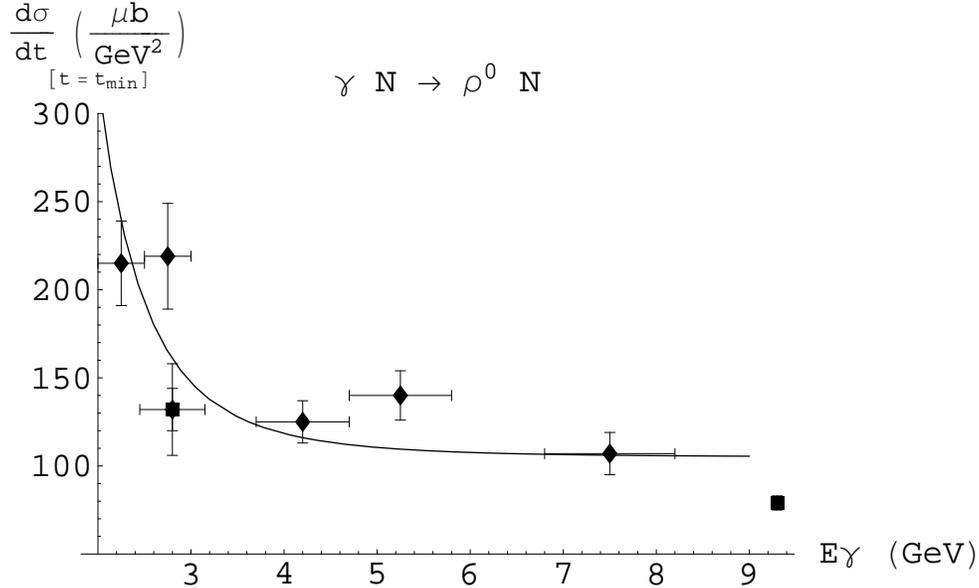}
\caption{We obtain this fit using data 
from Ref.~\cite{Ballam:1971wq,Eisenberg:1971ww,Ballam:1972eq}  (listed
in Ref.~\cite{Bauer:1977iq}).  We use an  inverse-fourth function at
low energies and apply a least-squares fit.  The peak in the parameterization yields a small
effect from Fermi motion (at a few GeVs) because of the small mass of the $\rho^{0}$.}
\label{fig:rhofit}
\end{figure}

First we consider the $\gamma N \rightarrow \rho^{0} N$ differential
cross section.  For this we use a simple exponential $t$-dependence
with a typical exponential slope of $B = 7.0$~GeV$^{-2}$ and an over
all normalization of $105$~$\mu b$/GeV$^{2}$.  (See,
e.g. Ref.~\cite{Bauer:1977iq}.)  It is known that at low energies the
normalization undergoes a steep rise.  We take this into account in
our calculation by including a factor of $(1 +
\frac{a}{E_{\gamma}^{4}})$ in the overall normalization and then doing
a least squares fit to obtain the parameter, $a$.  We find that $a \approx 32.7$.  The 
cross section is thus,
\begin{equation}
\frac{d \sigma}{dt} = 105 \frac{ {\mathrm \mu b }}{{\mathrm GeV^{2} }} \left( 1 + \frac{32.7 \, {\mathrm GeV}}{E_{\gamma}^{4}}  \right)  e^{7.0 \,{\mathrm GeV}^{-2} t}
\end{equation}
The result is shown
in Fig.~\ref{fig:rhofit}.  As is seen in the main part of the text, the
variation is too weak to introduce a very large effect on the final $\gamma
D \rightarrow \rho^{0} D$  cross section from Fermi motion.

\begin{figure}
\centering \includegraphics [scale=1.3]{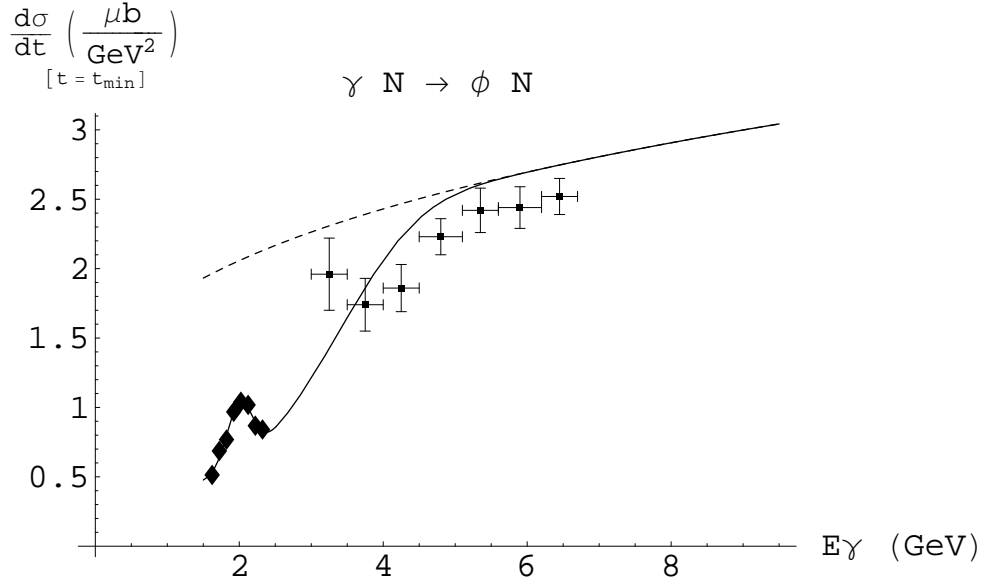}
\caption{The low energy data here is from Ref.~\cite{Mibe:2005er}.  
The curve at high energies was taken from Ref.~\cite{Bauer:1977iq}.
The dashed curve shows its extension to lower energies.  The  high
energy data is taken from Ref.~\cite{Behrend:1978ik}  and is presented
to establish the consistency of the high energy parameterization.  The
curve at low energies has been fit to the  low energy data of
Ref.~\cite{Mibe:2005er} using a least-squares fit.}
\label{fig:phifit}
\end{figure}

The case of the $\phi$-meson is more complicated due to the irregular
behavior near threshold.  The main point is to interpolate smoothly between recent 
low energy data and the standard higher energy parameterization.
The normalization of the low energy  data,
taken from recent experimental work in Ref.~\cite{Mibe:2005er}, is
obtained from an effective Pomeron and pseudo-scalar exchange
model~\cite{Titov:2003bk} as it was presented in Ref.~\cite{Mibe:2005er}.  We continue to use 
this so that our model will be consistent with current work. The high energy  parameterization was
obtained in Ref.~\cite{Bauer:1977iq} by fitting a diffraction-like
cross section to a large set of experimental data.  We want to
interpolate quickly but smoothly between the low energy and high
energy data.  There is an exponential factor, $e^{Bt}$, associated
with both the high and  the low energy behavior, but the slope, $B$,
is around $3.4$~GeV$^{-2}$ for the low energy behavior ($E_{\gamma}
\lesssim 4.0$~GeV) while it is around $4.8$~GeV$^{-2}$ for the  high
energy behavior.  Thus, for the exponential slope we use,
\begin{equation}
  B(E_{\gamma}) = \left( 4.8 - (4.8 - 3.4) e^{-0.001 \, {\mathrm GeV}^{-4}  E_{\gamma}^4} \right) \, {\mathrm GeV}^{-2}. \label{param1}
\end{equation}
Next, for the low energy region, there is no Regge slope.  That is,
$\alpha^{\prime} = 0$ in  the factor, $s^{\alpha^{\prime}t}$.  But, in
the high energy region, $\alpha^{\prime} = .27$~GeV$^{-2}$.  Thus, we
use,
\begin{equation}
  \alpha^{\prime}(E_{\gamma}) = .27 (1 - e^{-.001 \, {\mathrm GeV}^{-4} 
  E_{\gamma}^{4}}) \, {\mathrm GeV}^{-2}  . \label{param2}
\end{equation}
Now we consider the behavior of $\frac{d \sigma}{dt}_{\tilde{t} = 0}$ for photo-production of the $\phi$-meson 
from a proton target.  The
high energy parameterization in Ref.~\cite{Bauer:1977iq} is
\begin{equation}
\left. \frac{d \sigma}{dt} \right|_{t = 0} = 1.34 s^{.28},
\label{param3}
\end{equation}
and, as in Ref.~\cite{Bauer:1977iq},  over-all units will be understood to be $\frac{\mu b}{GeV^{2}}$.
In order to match to the  data of Ref.~\cite{Mibe:2005er} we want a
peak to appear at around $E_{\gamma} = 2.0$~GeV.  Therefore, we adjust
the parameterization to,
\begin{equation}
  \left. \frac{d \sigma}{dt} \right|_{\tilde{t} = 0} = 1.34 s^{.28} (1 + a e^{-b
  (E_{\gamma} - c)^2}). \label{param4}
\end{equation}
We use Eq.(\ref{param4}) to fit to the low energy data of
Ref.~\cite{Mibe:2005er} while assuring  that the high energy
parameterization of Eq.(\ref{param3}) is reproduced at high energies.
We find: $a = .71$, $b = 16.5$~GeV$^{-2}$, and $c = 2$~GeV$^{-2}$.
 Finally, we note that the low energy data is actually given for $t =
t_{min}$ rather than $t = 0$.   Therefore, we must be sure to include
a factor of $e^{B(E_{\gamma}) t_{min}}$ in the final result for
$ \left. \frac{d \sigma}{dt} \right|_{\tilde{t} = 0}$.  The result of our parameterization for
$ \left. \frac{d \sigma}{dt} \right|_{\tilde{t} = 0}$ for  the $\phi$-meson is shown in
Fig.~\ref{fig:phifit}.  We point out that in the intermediate energy range at around $E_{\gamma} = 3$~GeV,
the energy dependence is not completely flat, but it is smooth, and slow enough that its effect may 
be treated as a small correction.

\clearpage

\section{Deuteron Polarization}
\label{sec:app3}

In this appendix, we give an over view of the treatment of deuteron 
spin as it is presented in~\cite{Franco:1969qj}.
In order to evaluate the cross section, we must determine how the
operator, $\tilde{\Psi}^{\dagger}_{m^{\prime}}({\bf k} - \frac{{\bf
l}}{2}) \tilde{\Psi}_{m}( {\bf k} )$ acts on the spin-1 ground state
of the deuteron.  The non-relativistic deuteron wave function in
momentum space is written in terms of S and D states via the formula,
\begin{equation}
\tilde{\Psi}_{m}({\bf k}) = \left[ \tilde{u}(k) -  8^{-1/2}
\tilde{w}(k) \hat{S}_{12} \right] | \hat{q}, m \rangle \label{eq:pol1}
\end{equation} 
where,
\begin{equation}
\begin{split}
\tilde{u}(k) \equiv \frac{1}{\sqrt{2} \pi} \int_{0}^{\infty} r dr
j_{0}(kr) u(r)
 \\ \tilde{w}(k) \equiv 
\frac{1}{ \sqrt{2}\pi} \int_{0}^{\infty} r dr j_{2}(kr)
w(r). \label{eq:pol2}
\end{split}
\end{equation}
The real functions, $u(r)$ and $w(r)$, are taken from any realistic
model of the deuteron wave function, and in our computations we use
the Paris potential~\cite{Lacombe:1980dr}.  The functions, $j_{0}$ and $j_{2}$, are 
the zeroth and second order spherical Bessel functions.  In Eq.~(\ref{eq:pol1}), $|
\hat{q}, m \rangle$ is a spin-one spinor representing the total
angular  momentum of the deuteron, and $\hat{q}$ is the quantization axis.  The tensor operator, $\tilde{S}_{12}$
acts upon the  total angular momentum state to produce a sum over
total spin states.  In terms of the spins of the nucleons, it is given by:
\begin{equation}
\tilde{S}_{12} = \frac{3 \, ({\bf \sigma_{1}} \cdot {\bf r}) \, ({\bf \sigma_{2}} \cdot {\bf r}) - {\bf \sigma_{1}} \cdot {\bf \sigma_{2}} }{r^{2}}.
\end{equation}
The projection onto total spin states is,
\begin{eqnarray*}
  \hat{S} |0 , \hat{q} \rangle  & = & \sqrt{\frac{48 \pi}{5}}
  Y_{2}^{1}(\theta ,\phi ) | -1 \rangle  - \sqrt{\frac{64 \pi}{5}}
  Y_{2}^{0}(\theta ,\phi ) | 0 \rangle + \sqrt{\frac{48 \pi}{5}}
  Y_{2}^{-1}(\theta ,\phi ) | 1 \rangle
  \label{eq:pol3} \\ \hat{S} | -1 , \hat{q} \rangle & = & \sqrt{\frac{16
    \pi}{5}} Y_{2}^{0}(\theta ,\phi ) | -1 \rangle - \sqrt{\frac{48
  \pi}{5}} Y_{2}^{-1}(\theta ,\phi ) | 0 \rangle + \sqrt{\frac{96
  \pi}{5}} Y_{2}^{-2}(\theta ,\phi ) | 1 \rangle \label{eq:pol4} \\
  \hat{S}| 1 , \hat{q} \rangle & = & \sqrt{\frac{96 \pi}{5}}
  Y_{2}^{2}(\theta ,\phi ) | -1 \rangle - \sqrt{\frac{48 \pi}{5}}
  Y_{2}^{1}(\theta ,\phi ) | 0 \rangle + \sqrt{\frac{16 \pi}{5}}
  Y_{2}^{0}(\theta ,\phi ) | 1 \rangle . \label{eq:pol5}
\end{eqnarray*}
The functions, $Y$, are the usual spherical harmonic functions.
With these equations, we can calculate the effective form factor for
each polarization, and then sum and average over final/initial
deuteron polarizations.


\begin{thebibliography}{05}

\bibitem{Bauer:1977iq}
  T.~H.~Bauer, R.~D.~Spital, D.~R.~Yennie and F.~M.~Pipkin,
  Rev.\ Mod.\ Phys.\  {\bf 50}, 261 (1978)
  [Erratum-ibid.\  {\bf 51}, 407 (1979)].

\bibitem{feynman}R.P. ~Feynman ``Photon-Hadron 
Interactions" Addison Wesley Longman, Inc., 1972.

\bibitem{Franco:1965wi}
  V.~Franco and R.~J.~Glauber,
  Phys.\ Rev.\  {\bf 142}, 1195 (1966).

\bibitem{SLACexp}
 R.L.~Anderson {\it et al.}, 
 Phys.\ Rev.\ D {\bf 4}, 3245 (1971);
 I.D.~Overman, Ph.D. thesis,  SLAC-140, UC-34, 1971.

\bibitem{VM2}L.~Frankfurt, G.~Piller, M.~Sargsian and M.~Strikman, 
         Eur.\ Phys.\ J.\ A {\bf 2}, 301 (1998). \\
  L.~Frankfurt, W.~Koepf, J.~Mutzbauer, G.~Piller, M.~Sargsian and M.~Strikman,
  Nucl.\ Phys.\ A {\bf 622}, 511 (1997)
  [arXiv:hep-ph/9703399].

\bibitem{Titov:1999eu}
  A.~I.~Titov, T.~S.~Lee, H.~Toki and O.~Streltsova,
  Phys.\ Rev.\ C {\bf 60}, 035205 (1999).

\bibitem{Titov:2002zy}
  A.~I.~Titov, M.~Fujiwara and T.~S.~H.~Lee,
  Phys.\ Rev.\ C {\bf 66}, 022202 (2002)
  [arXiv:nucl-th/0207079].

\bibitem{Oh:2003gm}
  Y.~s.~Oh,
  J.\ Korean Phys.\ Soc.\  {\bf 43}, S20 (2003)
  [arXiv:nucl-th/0301011].

\bibitem{Frankfurt:1985cv}
  L.~L.~Frankfurt and M.~I.~Strikman,
  Nucl.\ Phys.\ B {\bf 250}, 143 (1985).

\bibitem{Kolbig:1968rm}
  K.~S.~Kolbig and B.~Margolis,
  Nucl.\ Phys.\ B {\bf 6}, 85 (1968).

\bibitem{Gribov:1968jf}
  V.~N.~Gribov,
  Sov.\ Phys.\ JETP {\bf 29}, 483 (1969)
  [Zh.\ Eksp.\ Teor.\ Fiz.\  {\bf 56}, 892 (1969)].

\bibitem{Bertocchi}L.~ Bertocchi, Nuovo Cimento, A {\bf 11} 45 (1972).

\bibitem{edepn}L.~L.~Frankfurt, W.~R.~Greenberg, G.~A.~Miller, 
               M.~M.~Sargsian and M.~I.~Strikman, 
               Z.\ Phys.\ A {\bf 352}, 97 (1995)

\bibitem{GEA}L.~L.~Frankfurt, M.~M.~Sargsian and M.~I.~Strikman,
         Phys.\ Rev.\ C {\bf 56}, 1124 (1997); 
         M.M.~Sargsian,  Int.\ J.\ Mod.\ Phys.\ E {\bf 10}, 405 (2001).

\bibitem{Jeschonnek:2000nh}
  S.~Jeschonnek,
  Phys.\ Rev.\ C {\bf 63}, 034609 (2001)
  [arXiv:nucl-th/0009086].

\bibitem{Mibe:2005er}
  T.~Mibe {\it et al.}  [LEPS Collaboration],
  arXiv:nucl-ex/0506015.

\bibitem{Alkhazov:1978et}
   G.~D.~Alkhazov, S.~L.~Belostotsky and A.~A.~Vorobev,
   Phys.\ Rept.\  {\bf 42}, 89 (1978).

\bibitem{Garrow}
K.~Garrow {\it et al.},
Phys.\ Rev.\ C {\bf 66}, 044613 (2002).

\bibitem{Binosi:2003yf}
  D.~Binosi and L.~Theussl,
  Comput.\ Phys.\ Commun.\  {\bf 161}, 76 (2004)
  [arXiv:hep-ph/0309015].


\bibitem{Brown:1973xt}G.E. ~Brown and A.D. ~Jackson ``The Nucleon-Nucleon 
Interaction" North-Holland Publishing Company, 1976.

\bibitem{Franco:1969qj}
  V.~Franco and R.~J.~Glauber,
  Phys.\ Rev.\ Lett.\  {\bf 22}, 370 (1969).

\bibitem{Lacombe:1980dr}
  M.~Lacombe, B.~Loiseau, J.~M.~Richard, R.~Vinh Mau, J.~Cote, P.~Pires and R.~De Tourreil,
  Phys.\ Rev.\ C {\bf 21}, 861 (1980).

\bibitem{Alvensleben:1971vy}
  H.~Alvensleben {\it et al.},
  Phys.\ Rev.\ Lett.\  {\bf 27}, 444 (1971).

\bibitem{PANIC}
  Talk given by Keito HOREI,
  ``Measurement of photoproduction of $\phi$-mesons near threshold by SPring-8/LEPS,''  
  Oct. 27, 2005 at the Particles and Nuclei International Conference in Santa Fe, NM - October 24-28, 2005
   To appear in the proceedings of PANIC05;
   http://panic05.lanl.gov/

\bibitem{Ishikawa:2005aw}
  T.~Ishikawa {\it et al.},
  Phys.\ Lett.\ B {\bf 608}, 215 (2005).

\bibitem{Titov:2003bk}
  A.~I.~Titov and T.~S.~H.~Lee,
  Phys.\ Rev.\ C {\bf 67}, 065205 (2003)
  [arXiv:nucl-th/0305002].

\bibitem{Ballam:1971wq}
  J.~Ballam {\it et al.},
  Phys.\ Rev.\ D {\bf 5}, 15 (1972).

\bibitem{Eisenberg:1971ww}
  Y.~Eisenberg, B.~Haber, E.~Kogan, E.~E.~Ronat, A.~Shapira and G.~Yekutieli,
  Nucl.\ Phys.\ B {\bf 42}, 349 (1972).

\bibitem{Ballam:1972eq}
  J.~Ballam {\it et al.},
  Phys.\ Rev.\ D {\bf 7}, 3150 (1973).

\bibitem{Behrend:1978ik}
  H.~J.~Behrend {\it et al.},
  Nucl.\ Phys.\ B {\bf 144}, 22 (1978).

\end{thebibliography}
\end{document}